\RequirePackage{fix-cm}
\documentclass[smallextended]{svjour3}       
\smartqed  
\usepackage{graphicx}
\usepackage{amsmath}
\usepackage{amsfonts}
\usepackage{mathtools}
\usepackage{natbib}
\usepackage{subfigure}
\usepackage{epsfig}
\usepackage{booktabs}
\usepackage{multirow}
\usepackage[linesnumbered,ruled,vlined]{algorithm2e}
\usepackage{paralist}
\usepackage{xcolor}
\usepackage{mwe}

\usepackage{graphicx}
\usepackage{blindtext}

\renewcommand{\cite}{\citep}

\DeclareMathOperator*{\minimize}{minimize \;}
\DeclareMathOperator*{\optimize}{optimize \;}

\begin{document}

\title{Variable Functioning and Its Application to Large Scale Steel Frame Design Optimization\thanks{This research was supported in part by the National Science Foundation (NSF) under Cooperative Agreement DBI-0939454. Any opinions, findings, and conclusions or recommendations expressed in this material are those of the author and do not necessarily reflect the views of the NSF.}
}


\author{Amir~H~Gandomi \and
Kalyanmoy~Deb \and
Ronald~C~Averill \and
Shahryar~Rahnamayan \and
and~Mohammad~Nabi~Omidvar
}


\institute{Amir H Gandomi (corresponding author) \at Faculty of Engineering and Information Technology, University of Technology Sydney, Australia \email{gandomi@uts.edu.au}
\and
Kalyanmoy Deb \at Department of Electrical and Computer Engineering, Michigan State University, East Lansing, MI 48824, USA \email{kdeb@egr.msu.edu}
\and
Ronald C Averill \at Department of Mechanical Engineering, Michigan State University, East Lansing, MI 48824, USA \email{averillr@egr.msu.edu}
\and
Shahryar Rahnamayan \at Department of Electrical, Computer and Software Engineering, University of Ontario Institute of Technology, Oshawa, ON, Canada \email{shahryar.rahnamayan@uoit.ca}
\and
Mohammad Nabi Omidvar \at School of Computing and Leeds University Business School, University of Leeds, UK \email{m.n.omidvar@leeds.ac.uk}
}

\date{Received: date / Accepted: date}

\maketitle

%

\begin{abstract}
To solve complex real-world problems, heuristics and concept-based approaches can be used in order to incorporate information into the problem. 
In this study, a concept-based approach called variable functioning ($Fx$) is introduced to reduce the optimization variables and narrow down the search space. In this method, the relationships among one or more subset of variables are defined with functions using information prior to optimization; thus, instead of modifying the variables in the search process, the function variables are optimized. By using problem structure analysis technique and engineering expert knowledge, the $Fx$ method is used to enhance the steel frame design optimization process as a complex real-world problem. The proposed approach is coupled with particle swarm optimization and differential evolution algorithms and used for three case studies. The algorithms are applied to optimize the case studies by considering the relationships among column cross-section areas. The results show that $Fx$ can significantly improve both the convergence rate and the final design of a frame structure, even if it is only used for seeding.

\end{abstract}

\section{Introduction}
\label{sec:introduction}

Real-world optimization problems are often complex and difficult to solve due to factors such as dimensionality, nonlinearity, and existence of complex constraints. To deal with this level of complexity, many approaches such as dimensionality reduction, approximation, and problem decomposition are common. Another way of dealing with complex problems is to turn to expert knowledge and incorporate it into the optimization process. For example, in the case of problem decomposition, knowledge of problem structure is needed, which might be known \emph{a priori}, or might be discovered automatically with relevant analysis techniques~\cite{mei2016competitive}.

Recently, gray-box optimization has been coined to refer to the optimization process of problems for which the structure is known~\cite{santana2017gray}. This is in contrast to black-box optimization, where zero knowledge of the problem is assumed. Although this might be a reasonable assumption for simulation type problems, its is not a realistic assumption for a wide range of optimization problems. For example, the use of a deterministic crossover operator which respects problem structure allows optimization of cast scheduling problems with up to a billion variables~\cite{deb2017population}. Similar approaches have been used successfully with Traveling Salesman Problems~\cite{whitley2010hybrid} and pseudo-boolean problems~\cite{tintos2015partition}.

In the context of evolutionary algorithms, two general approaches have been suggested to adapt an algorithm to the known characteristics of a given problem~\cite{de1988learning}:
\begin{inparaenum}
    \item To change the representation of the problem such that the traditional variation operators remain applicable;
    \item To devise new variation operators to work with the original representation of the problem perceived as ``natural''.
\end{inparaenum}
In this paper, we use the former approach to solve steel frame design optimization problems.

Frame design is one of the most popular optimization problems in structural engineering~\cite{saka2007optimum}. It is considered to be a complex problem due to the involvement of elaborate finite element models and the existence of several mechanical and geometric constraints relating to maximum and minimum stress, buckling, story and roof drifts. Furthermore, the existence of discrete or mixed-type variables, due to the need for predefined cross-sections in the frame construction industry, also adds to its complexity.

Due to its wide practical applications and versatility, an effective method of solving such problems can significantly reduce the construction cost. As a result, many researchers have attempted to optimize frame structure design as a complex, discrete problem, using a variety of methods~\cite{lamberti2011metaheuristic} including non-deterministic and stochastic algorithms~\cite{hasanccebi2010comparison, azad2015computationally}. The objective of frame design is to minimize the frame weight (relating to cost) subject to complex nonlinear constraints. For the steel frame structures, the design variables are usually the cross-sections of the beams and columns. In practice, these members must be chosen from a standardized set of cross-sections, which makes the problem discrete. 

In this paper, we propose a method called variable functioning ($Fx$) to change the problem representation with the aim of controlling its complexity and taming the curse of dimensionality by reducing the number of decision variables. In this method, the structural information of the problem is extracted using state-of-the-art variable interaction analysis methods based on which a functional mapping is created to map the input space to a lower dimensional space, based on the identified interaction patterns of original input variables. For the purposes of this paper, we adopted differential grouping (DG2)~\cite{omidvar2017dg2} to find the problem structure in the form of nonlinear relationships among the decision variables. These information can be visualized with heat-maps and/or variable interaction graphs to give further insight about the nature of the problem. The variable interaction information contains interesting patterns which are otherwise hard to discover even by the experts. These patterns are then used by the expert to devise a functional mapping that transforms the original complex problem into a simpler problem with fewer decision variables. Once the problem is reformulated, the resulting transformed problem can be optimized using any suitable optimizer.

Additional information can be embedded in an optimization process, in a heuristic way, to simplify or improve the process.
Moreover, the heuristics can be very effective if they are applied correctly.
There are some existing heuristics in the frame design optimization process, such as considering symmetry of a problem to decrease the number of variables and reduce the number of finite elements in the model~\cite{talatahari2015optimum}.
Additionally, fabrication conditions should be imposed on the construction of structural elements, such as requiring the same beam/column cross-section to be used for $N$ consecutive stories, resulting in a reduction in the number of problem variables~\cite{talatahari2015optimum}. 
However, unlike variable functions proposed in this paper, these methods are problem specific and cannot be generalized to a wider range of problems. Variable functioning, on the other hand, is based on automatic variable interaction analysis, which can be applied to a wide range of problems. It is worth mentioning that the proposed method is compatible with these ad-hoc techniques and can be considered as a complement.


The proposed approach is simple and can be coupled with any optimization algorithm; here, it is combined with particle swarm optimization (PSO) and differential evolution (DE) algorithms.
The proposed method is explained along with an illustrative example, and then applied to three steel frame design optimization problems.
The results show that the proposed approach can significantly improve both the convergence rate and the final solution of frame design optimization problems, even if it is only used in the initialization step, and not through entire search optimization process.  

The rest of this study is organized as follows: Section~\ref{sec:background} presents the formulation of the steel frame design problem along with the optimization algorithms and constraint handling schemes used in this study. The proposed variable functioning approach and its application to steel frame design problem are explained in Section~\ref{sec:fx} by means of an illustrative example. Section~\ref{sec:cases} is devoted to the three case studies used for benchmarking in this study. Finally, conclusions, discussions, and future insights are presented in Section~\ref{sec:conclusions}.


\section{Steel Frame Design Optimization Process}
\label{sec:background}

\subsection{Problem Formulation}

The main variables in a steel frame design optimization problem are usually member sections, which are grouped based on the fabrication conditions and symmetry of the structure.
Therefore, the optimization variables in a steel frame structure are cross-sections of each group, as $\vec{x} = (x_1, \dots, x_{n_\mathrm{g}})$ where $n_{\mathrm{g}}$ is the number of member groups.
As it is mentioned, the objective of frame design problems is usually the minimization of the frame weight ($W$) which can be formulated as:
\begin{equation}
    \minimize_{\vec{x} \in \Omega} W(\vec{x}) = \sum_{i=1}^{n_{\mathrm{g}}}{\rho \left( \sum_{j=1}^{n_{\mathrm{m}}}{L_{i,j}} \right)} x_i,
\end{equation}
where $\rho$ is the material density, $L_{i,j}$ is the length of $j$th element in the $i$th group, $ng$ is the number of groups, $n_{\mathrm{m}}$ is the number of members in the $i$th group, and $\Omega$ is the search space for the variables (cross-sections). Note that although other variables can be considered for steel frame design optimization, this study only considers cross-sections as design variables. Even though in the present work weight of the steel frames is considered as the objective function, a more detailed optimization approach could take into account the cost-efficiency of the design examples. Despite the fact that the main concern of the current study is reducing the search space, it can also be readily applied to problems where minimizing the cost is considered as the objective~\cite{pavlovvcivc2004cost}. In practice, cross-sections must be chosen from a predefined and standardized set of cross-sections (e.g. W-shapes), which makes the problem discrete. This problem is generally subject to stress constraints, maximum lateral displacement, and inter-story displacement constraints~\cite{talatahari2015optimum}. The stress constraints can be formulated as follows:
\begin{equation}
    v_{i}^{\sigma} = \left| \frac{\sigma_i}{\sigma_i^a} \right| - 1 \ge 0, \forall i \in \{1,\dots,n_{\mathrm{m}}\}
\end{equation}
where $\sigma_i$ and $\sigma_i^a$ are respectively the maximum stress and the allowable stress in the $i$th member.  Therefore, the number of stress constraints is equal to a number of members (elements) in the problem.
The maximum lateral displacement and inter-story displacement constraints can be respectively formulated as follows:
\begin{align}
v^{\Delta} &= \frac{\Delta_T}{H} - R \ge 0, \\
v^d_j &= \frac{d_j}{h_j} - R_I \ge 0,
\end{align}
where $\Delta_T$ is the maximum lateral displacement; $H$ is the height of the frame structure; $R$ is the maximum drift index; $d_j$ is the inter-story drift; $h_j$ is the height of the $j$th floor; $ns$ is the total number of stories; and $R_I$ is the inter-story drift index permitted by the standard design code in engineering practice.
The allowed inter-story drift index is taken as 1/300, based on the American Institute of Steel Construction (AISC) design code (AISC 2011). For the Load and Resistance Factor Design (LRFD) interaction formula constraints (AISC 2001, Equation H1-1a,b) are formulated as follow:
\begin{equation}
v_i^I \!=\! 
\begin{dcases}
    \frac{P_{\mathrm{u}}}{2\phi_c P_{\mathrm{n}}} \!+\!\! \left(\frac{M_{\mathrm{ux}}}{\phi_b M_{\mathrm{nx}}} \!+\!\! \frac{M_{\mathrm{uy}}}{\phi_b M_{\mathrm{ny}}} \right) \!-\!\! 1 \le 0, & \frac{P_{\mathrm{u}}}{\phi_c P_{\mathrm{n}}} < 0.2 \\
    \frac{P_{\mathrm{u}}}{\phi_c P_{\mathrm{n}}} \!+\!\! \frac{8}{9}\left( \frac{M_{\mathrm{ux}}}{\phi_b M_{\mathrm{nx}}} \!+\!\! \frac{M_{\mathrm{uy}}}{\phi_b M_{\mathrm{ny}}} \right) \!-\!\! 1 \le 0, & \frac{P_{\mathrm{u}}}{\phi_c P_{\mathrm{n}}} \ge 0.2,
\end{dcases}
\end{equation}
where $P_{\mathrm{u}}$ is the required strength (tension or compression); $P_{\mathrm{n}}$ is the nominal axial strength (tension or compression); $\phi_c$ is the resistance factor ($\phi_c=0.9$ for tension, $\phi_c =0.85$ for compression); $M_{\mathrm{ux}}$ and $M_{\mathrm{uy}}$ are the required flexural strengths in the $x$ and $y$ directions, respectively; $M_{\mathrm{nx}}$ and $M_{\mathrm{ny}}$ are the nominal flexural strengths in the $x$ and $y$ directions (for two-dimensional structures, $M_{\mathrm{ny}}=0$); and $\phi_b$ is the flexural resistance reduction factor ($\phi_b=0.9$).
The effective length factors of members ($K$) are required to compute the allowable compression and Euler buckling stresses.

\subsection{Algorithms}
Metaheuristics are the global and stochastic optimization algorithms, and they are generally inspired by nature.
Based on the sources of information, metaheuristics can be divided into two classes: swarm intelligence~\cite{slowik2018nature}, the algorithms mimic a swarm behavior, and evolutionary computation~\cite{eiben2015evolutionary} which use evolutionary mechanism such as crossover, mutation, and selection.
Both classes of metaheuristics have been widely used for simulation optimization (such as FE analysis). Also, they have been successfully applied in complex frame design optimization problems~\cite{saka2007optimum, gholizadeh2016seismic, ghasemi2011ant}. 

In this study, a particle swarm optimization (PSO) algorithm is used as a classical swarm intelligence and a differential evolution (DE) algorithm is used for the frame optimization as an acclaimed evolutionary algorithm to couple with the proposed variable functioning approach. It should be noted that finding the best algorithm(s) is not the purpose of this study and these two algorithms have been used to represent two classes of metaheuristics. 

 The PSO algorithm mimics the social behavior of bird flocks and fish schools, and it was initially suggested by~\citet{kennedy1995j}.
 This algorithm is one of the best-established swarm intelligence algorithms, and has been applied to many structural optimization problems.
 PSO algorithm is population-based, and particles forage the search space to find the best solutions.
 DE~\citet{storn1997differential} is a population-based evolutionary optimization algorithm which uses three operators (selection, mutation, and crossover) to lead the solution toward the global optimum. PSO and DE algorithms have been used in this study for steel frame design optimization.
Note that this study does not propose a new optimization algorithm. The aim of this study is to show how the proposed method work with global optimization algorithms in order to reduce the search space, more specifically the design space of steel frame design.

\subsection{Constraint Handling}
The constraint handling is based on feasibility rules proposed by~\citet{deb2000efficient} as follows:
\begin{enumerate}
\item If both solutions are feasible, the one with the better objective function value is preferred.
\item A feasible solution is preferred to an infeasible one.
\item If both solutions are infeasible, the one with smaller amounts of constraint violation is preferable.
    The amount of constraint violation is also normalized according to~\citet{becerra2006cultured}:
\end{enumerate}

\begin{equation}
    G(\vec{x}) = \sum_{i=1}^{n_{\mathrm{c}}}{\frac{g_i(\vec{x})}{g_{\mathrm{max},i}}},
\end{equation}
where $n_{\mathrm{c}}$ is the number of constraints, $g_i(\vec{x})$ are the $i$th constraint of the problem, and $g_{\mathrm{max}\;i}$ is the largest violation of the $i$th constraint so far.
Based on the rules, a fitness function is proposed by~\citet{deb2000efficient} in order to penalize the solutions that violate the constraint(s). The penalized fitness function is formulated as follows:

\begin{equation}
f_{\mathrm{p}}(\vec{x}) = 
\begin{dcases}
f(\vec{x}) & \text{if } G(\vec{x}) \ge 0 \\
f_{\mathrm{max}} + G(\vec{x}) & \text{otherwise},
\end{dcases}
\end{equation}
where $f_{\mathrm{max}}$ is the objective function value of the worst feasible solution in the population.
Using this strategy, infeasible solutions are only compared based on their normalized constraint violation. Note that other constraint handling schemes, such as Automatic Dynamic Penalisation method (e.g. \citet{montemurro2013automatic}), may perform better, which need to be evaluated since it depends on the optimization algorithm, problem, as well as the implementation details of the proposed approach. In both DE and PSO algorithms, solutions may go out of the defined boundaries.
In DE, if a component of a solution violates either upper or lower bounds, it is returned to the violated bound.
In PSO, however, if the solution is returned to the boundary, it has a high probability of violating the boundary again.
This is because the inertia of the previous motion contributes to the current motion.
Therefore, when a particle component violates a boundary, its related velocity component ($\mathbf{V}$) is reversed to return the solution to the boundary and also overcome the problem.
This boundary constrained handling can be formulated as follows:

\begin{equation}
v^{(t+1)}_{i,j} =
\begin{dcases}
v^{(t+1)}_{i,j} & \text{feasible} \\
-v^{(t+1)}_{i,j} & \text{infeasible}.
\end{dcases}
\end{equation}


\section{Variable Functioning}
\label{sec:fx}

\subsection{Methodology}
Most real-world problems are complex and sometimes they are dealing with black/gray box models.
Optimization of such problems is also difficult because we do not have much information from the systems.
In such cases embedding information and knowledge can be very helpful and boost the optimization process.
This information can be obtained from different sources such as statistical test, engineering point of view, expert systems, etc.


Such knowledge can be utilized as additional information even before search process, and their proper adaptation can significantly improve the optimization process.
In this section, an approach is proposed and explained for incorporating information from different sources in order to narrowing down the search space and potentially reducing the problem dimension. Next section, particularly focus on applying the proposed approach on steel frame design optimization problem. 

Here engineering domain knowledge is embedded into the steel frame optimization process, which can be used in similar problems, as well.

In an optimization problem, one (or more) objective function(s) should be optimized with respect to some variable vector $\vec{x} = (x_1, \dots, x_n)$, which can be simply formulated as:
\begin{equation}
\underset{\vec{x} \in \Omega}{\text{Optimize }} f(x_1, \dots, x_n).
\end{equation}

Considering the relationship among a set of variables, a general relationship could be defined using a mathematical function $\zeta: \mathbb{R}^m \to \mathbb{R}^q$. For example, if the first $q$ variables, $(x_1,\dots,x_q)$, have a functional relationship that can be expressed with $\zeta$, they can be replaced with a function with $m$ variables $\vec{y} = (y_1,\dots,y_m)$. Now, the optimization problem can be represented as: 
\begin{equation}
    \optimize_{\vec{x} \in \Omega_{\vec{x}} \wedge \vec{y} \in \Omega_{\vec{y}}}  f(\zeta(\mathbf{y}), x_{q+1}, \dots, x_n),
\end{equation}
or simply
\begin{equation}
    \optimize_{\vec{x} \in \Omega_{\vec{x}} \wedge \vec{y} \in \Omega_{\vec{y}}} \hat{f}(y_1, \dots, y_m, x_{q+1}, \dots, x_n).
\end{equation}
where $m$ is the number of variables in the functioning relationship ($\zeta$).
Using the new formulation ($\hat{f}$), the optimization results always satisfy the defined relationship for the first $q$ variables. Now, if $m < q$, it means the problem dimension is also reduced, which can simplify the problem by reducing the search space.  
If more than one set of variables has the functioning relationships, the generalized form of the approach for any number of functions can be formulated as follow:
\begin{align}
    \underset{\vec{x} \in \Omega_{\vec{x}} \wedge \vec{y} \in \Omega_{\vec{y}}}{\text{optimize }} 
    f(&\zeta_1(\mathbf{y}_1),\dots, \zeta_s(\mathbf{y}_s),  x_{q+1}, \dots, x_n),
%
\end{align}
where $s$ is the number of functioning relationships and each functioning relationship $\zeta_i(\mathbf{y}_i): \mathbb{R}^{m_i} \to \mathbb{R}^{q_i}$.
From the general formulation, if $\sum_{j=1}^{s}{m_j} < \sum_{j=1}^{s}{q_j}$, it means that the problem dimension is reduced, which is equal to their difference, $\sum_{j=1}^{s}{(q_j - m_j)}$.
The proposed approach can be coupled with any generalized optimization algorithm, as it only alters optimization problem formulation.

 Several approaches are proposed recently in order to group decision variables of a black-box problem, such as adaptive monotonicity checking~\cite{munetomo1999identifying, chen2010large}, variable partitioning~\cite{ray2009cooperative}, and min/max variance decomposition~\cite{liu2013scaling}. However, all these methods still have a low grouping accuracy. This drawback results in several recent studies, including variable interaction learning~\cite{chen2010large}, statistical learning decomposition~\cite{sun2012cooperative}, and meta-modeling decomposition~\cite{mahdavi2014cooperative}.  
 In one of the successful studies, \citet{omidvar2014cooperative} proposed the differential grouping (DG) to determine the nonseparable groups and later on they extended it as DG2~\cite{omidvar2017dg2}. DG2 has shown superior performance concerning grouping accuracy methods and it does not have any parameter to tune~\cite{omidvar2017dg2}.

\subsection{Variable Functioning for Steel Frame Structures}

\begin{figure}[t]
\begin{minipage}{0.45\columnwidth}
\centering
\includegraphics[width=1.1\columnwidth]{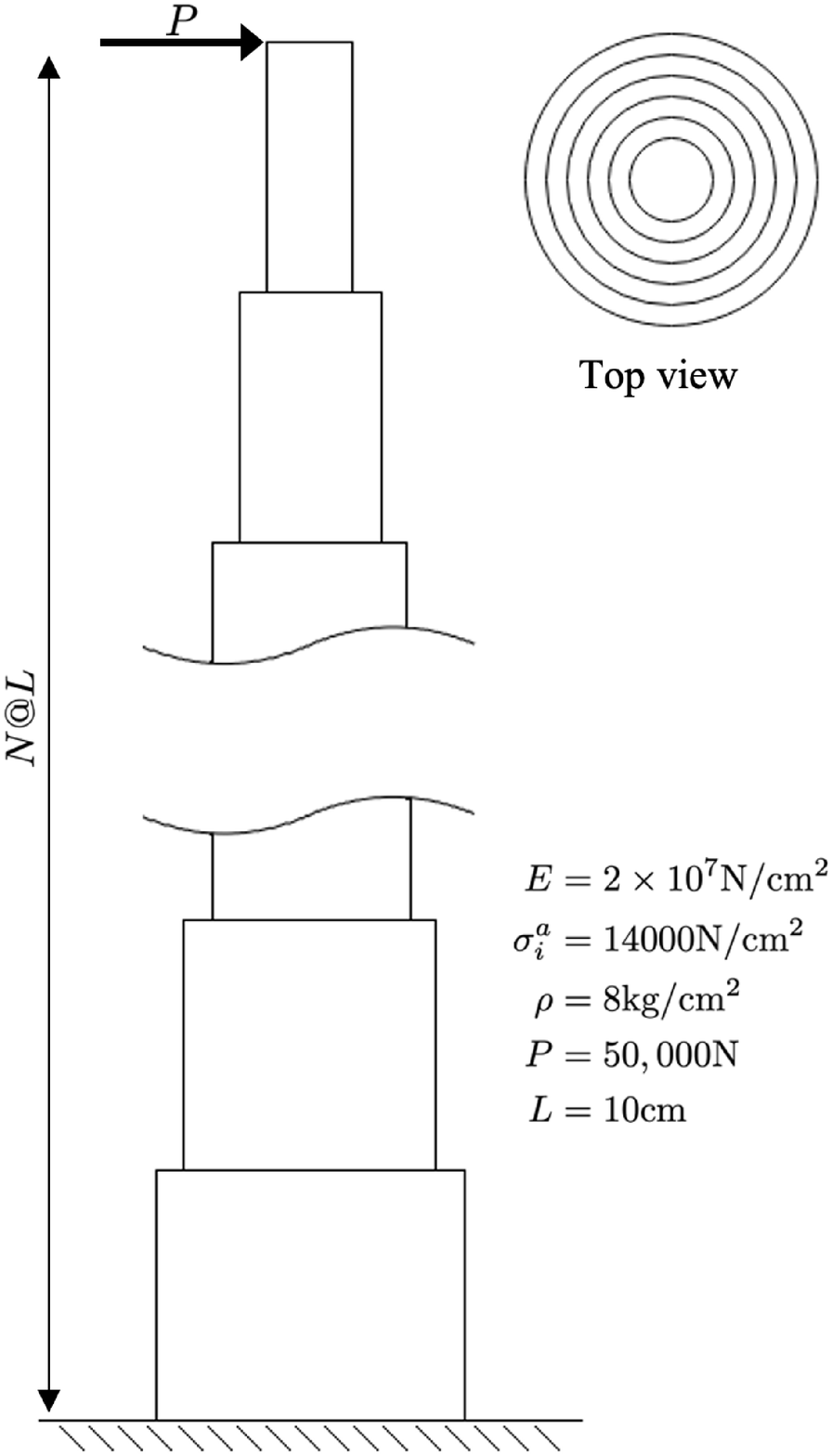}
\caption{Stepped column design problem with circular cross-sections under lateral load.}
\label{fig:sbeamc}
\end{minipage}%
\hfill
\begin{minipage}{0.45\columnwidth}
\centering
\includegraphics[width=1\linewidth]{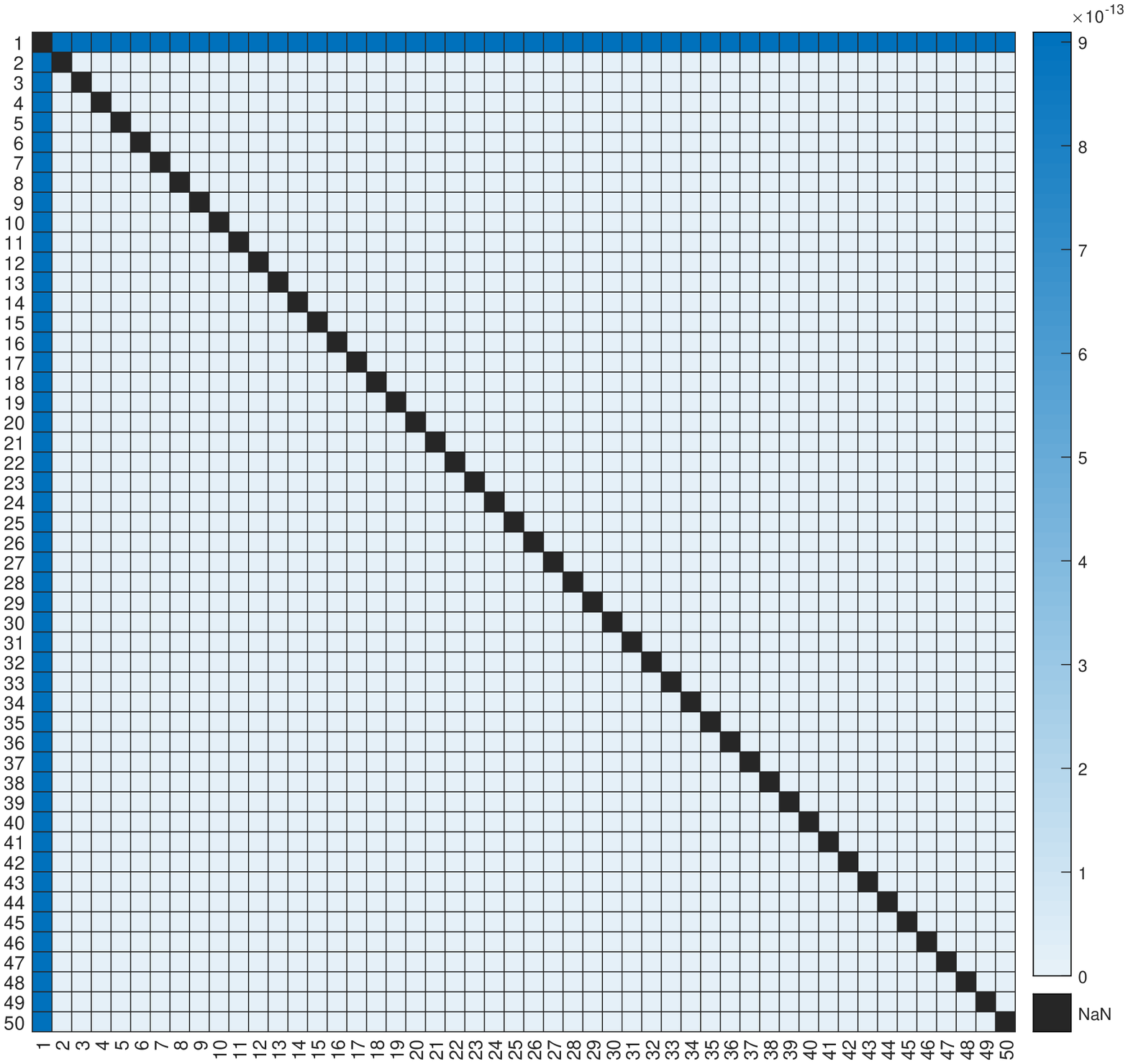}
\caption{Interaction matrix.}
\vspace{3mm}
\label{fig:dg50seg}
\centering
\includegraphics[width=1\linewidth]{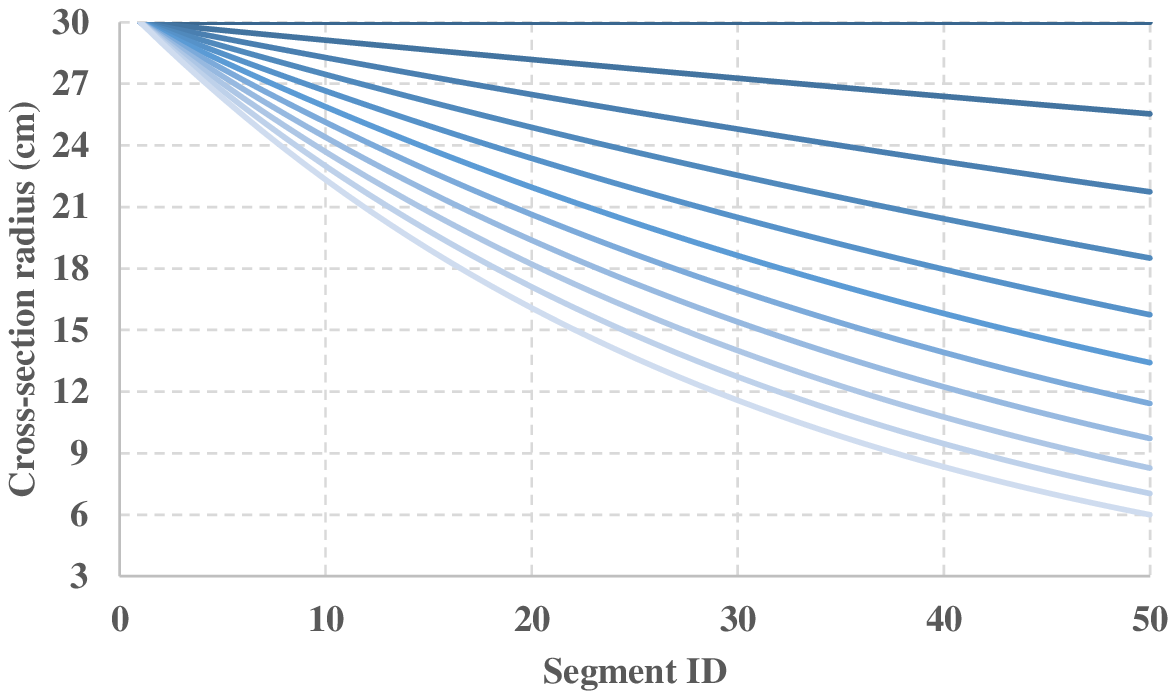}
\caption{Sample solutions for the cross-section of the base column equal to 30 cm.}
\label{fig:example}
\end{minipage}
\end{figure}

Structural engineering knowledge and concepts are usually used for the formulation of structural optimization problems.
For instance, to consider the fabrication conditions in steel frame optimization problems, the beam and column sections are grouped into two/three consecutive stories.
In this section, we aim to find ways to embed information and knowledge to find relationship between column cross-sections as main variable in steel frame design optimization.
A single stepped column under a lateral load is shown in Figure~\ref{fig:sbeamc}.
Minimization of the column weight is the objective, subject to satisfying the maximum stress constraint in each segment, and variables are the radii of the cross-sections. 



One way of finding variable interactions in a black box model is performing the differential grouping test~\cite{omidvar2014cooperative}.
This method is recently improved by~\cite{omidvar2017dg2}, named DG2.
Here, DG2 method is used to find interaction between variables and its results for a column with 50 segments is visualized in Figure~\ref{fig:dg50seg}.
From this heat map, it is clear that variables have high dependencies to the first variable (cross-section of the base segment).


From an engineering point of view, there are several reasons why the cross-section of a floor should be less than or equal to a cross-section in the lower floor.
First it is because of the physics of the problem, and the fact that each column cross-section carries the cumulative axial loads above it.
Additionally, installing a cross-section on a smaller cross-section in a column is not practical, and also notably difficult, as it cannot transfer the load correctly.
Therefore, the cross-section area should be monotonically decreased as the story number increases.
In other words, the story height and the column cross-section area have an inverse relationship. 

Considering the two mentioned points: 
\begin{itemize}
\item dependencies of all cross sections to the first one (based on the DG2 test results illustrated in Figure~\ref{fig:dg50seg}); and
\item monotonically decreasing column cross-section area with increasing heights (based on engineering points of view);
\end{itemize}
the following function is defined to relate the cross-section area of a column by means of their heights, as:
\begin{equation}
A(h) = \frac{A(0)}{\alpha^h}
\label{eq:Ah}
\end{equation}
where $A(0)$ is the cross-section area of the base segment; $\alpha$ is a new variable; $h$ is height of a cross-section from the base (or ground); and $A(h)$ is the cross-section area of the column located at the height of $h$.
For the base cross-section, $h=0$, $\alpha = 1$ and from~\eqref{eq:Ah}, $A(h) = A(0)$.
In this equation, all the variables (cross-section areas) are related to the first variable by the new variable, $\alpha$, and cross-section height and therefore, all a column' cross-section areas can relate to each other using a function with just two variables: the cross-section area of the base column, $A(0)$, and $\alpha$.
To monotonically reduce the cress-section areas while increasing the height, $\alpha$ should be greater than 1.
Furthermore, the largest value of $\alpha$ is for the time when the lowest possible cross-section area is located at the end of the column (highest cross-section), which can be calculated from the following equation: 
\begin{equation}
\alpha_{\mathrm{max}} = \sqrt[h_{\mathrm{u}}]{\frac{A_{\mathrm{max}}}{A_{\mathrm{min}}}}, 
\end{equation}
where $h_{\mathrm{u}}$ is the height of the uppermost column section from the ground. $A_{\mathrm{min}}$ and $A_{\mathrm{max}}$ are the smallest and the largest cross-section areas among the predefined sections, respectively.
Based on the above discussion, the boundaries of the new variable are $\Omega_{\alpha}: \alpha \in [1, \alpha_{\mathrm{max}}]$.
Using this approach, all the variables (cross-section areas) above the base are replaced with $\alpha$ in the new formulation. It should be noted that this function can be used for different sets of columns. In other words, a problem with $s$ columns and $n$ cross-sections, $f(A_1, \dots, A_n)$, is replaced with $f(A^1(h), \dots, A^s(h), A_{i+1}, \dots, A_n)$ after defining $s$ functions.
Note that other rules may apply for other problems like weak beam-strong column rule for the seismic regions that are not low-risk.

\subsection{Illustrative Example}

A simple example for applying the proposed approach is a single stepped column under a lateral load, which is created here to illustrate the proposed method.
Here, minimizing the column weight is the objective, subject to satisfying the maximum stress constraint in each segment, and variables are the radii of the sections (here radii are used instead of area).
Then, the optimization problem is formulated as follows:
\begin{equation}
    \underset{r \in \Omega_r}{\text{minimize }} f(\vec{r}) = \rho L \sum_{i=1}^{N}{\pi r_i^2},
\end{equation}
where $N$ is the number of segments which is equal to 50 here; $\vec{r} \in \{r_1, \dots, r_N\}$ and $r_i$ is the radius of the $i$th segment subject to $r \in [3,50]^{50} cm$; $\rho$ is the density, $L$ is the length of each segment.
Subject to:
\begin{equation}
g(r) = \sigma_i - \sigma^a_i \le 0,
\end{equation}
where $\sigma_i$ is the maximum bending stress at the bottom of the $i$th segment, and $\sigma^a_i$ is the allowable stress. 
To implement the proposed variable functioning strategy, the variables are converted as:
\[
\{r_1,\dots,r_N\} \mapsto \{r_1, \alpha\}
\]
The height vector of the cross-section is now equal to:
\[
    h(i) = (i-1)L
\]

The $h_\mathrm{u}$ here is $h_{\mathrm{max}} = (N-1)L=49L=490\mathrm{cm}$ and, therefore, the $\alpha$ boundary can be obtained as:
\[
\alpha_{\mathrm{max}} \sqrt[490]{\frac{50}{3}} = 1.0057581
\]

And, the $\alpha$ bounds should be as: $\Omega_{\alpha}: \alpha \in [1, 1.00575817]$
All the segments' radii are related to the radius of the first segment and the $\alpha$ value.
For this problem, ten solutions with $r_1=30\mathrm{cm}$ are presented in Figure~\ref{fig:example}.
The straight line at the top shows the $\alpha=1$ and the lowest line (brightest line) shows the solution with $\alpha$ equal to $\alpha_{\mathrm{max}}$.
In this case, all the possible solutions will be placed within these two lines.
Based on the original formulation, each cross-section radius should be within 3 and 50cm.
Since $r_1=30\mathrm{cm}$, all radii should be less than 30cm. 
Also, components of a solution have an order (shown as lines in Figure~\ref{fig:example}) and are not distributed in the whole search space.
Therefore, it is clear how much this approach narrows down the search space, which can enhance the search process.


After reformulating the current problem, the new objective function can be redefined as follows:
\begin{equation*}
\begin{aligned}
& \underset{r_1 \in \Omega_{r} \wedge \alpha \in \Omega_{\alpha}}{\text{minimize}}
& & f(r_1, \alpha) = \rho L \sum_{i=1}^{N}{\pi \left( \frac{r_1}{\alpha^{h(i)}} \right)^2} \\
& \text{subject to}
& & g(r_1, \alpha) = \sigma_i-\sigma^a_i \le 0.
\end{aligned}
\end{equation*}

\begin{figure}[t]
    \centering
    \mbox{\subfigure[]{\psfig{figure=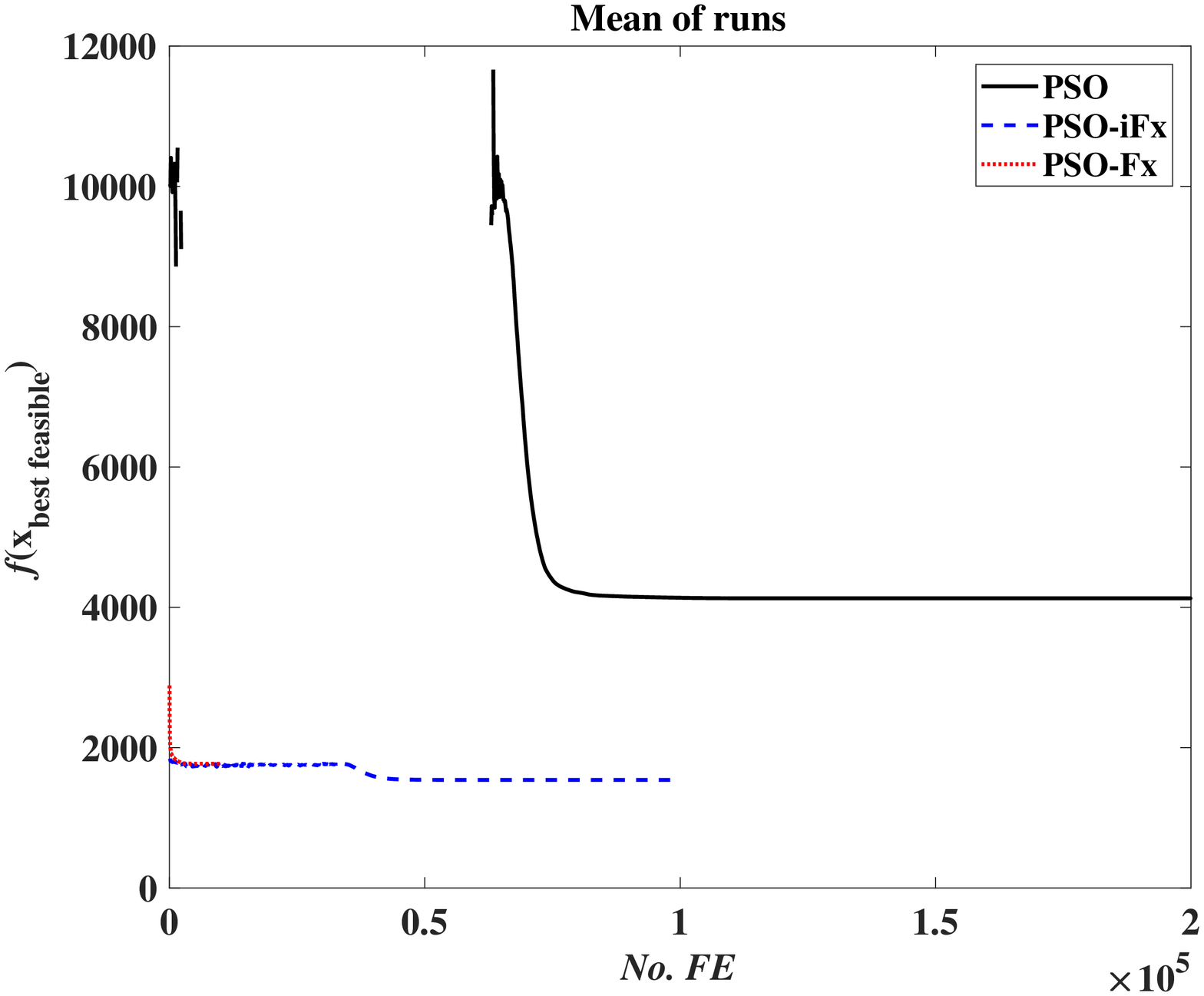,width=4.3cm}\label{sfig:pso}}\;
          \subfigure[]{\psfig{figure=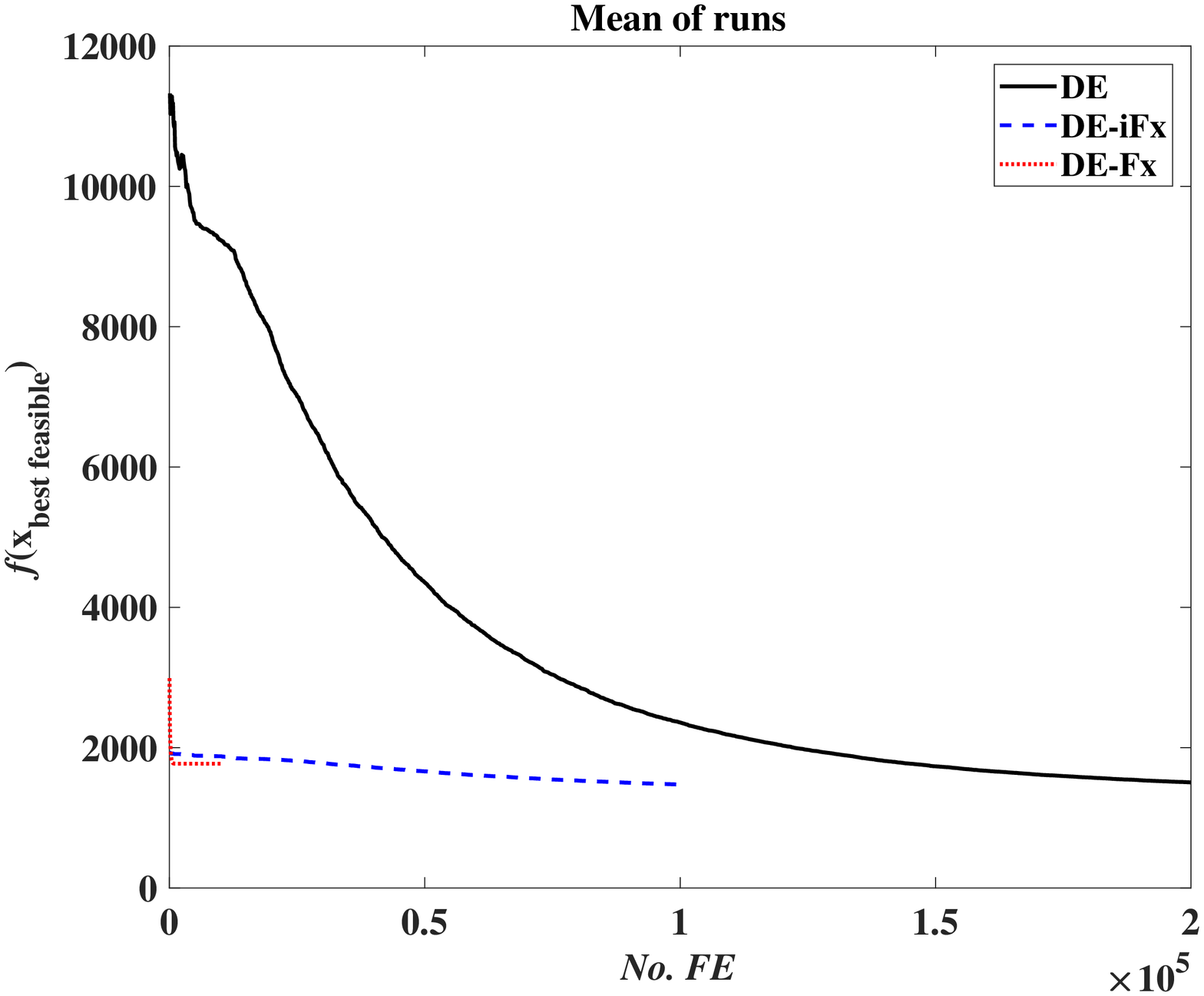,width=4.3cm}\label{sfig:de}}\;}
    \mbox{\subfigure[]{\psfig{figure=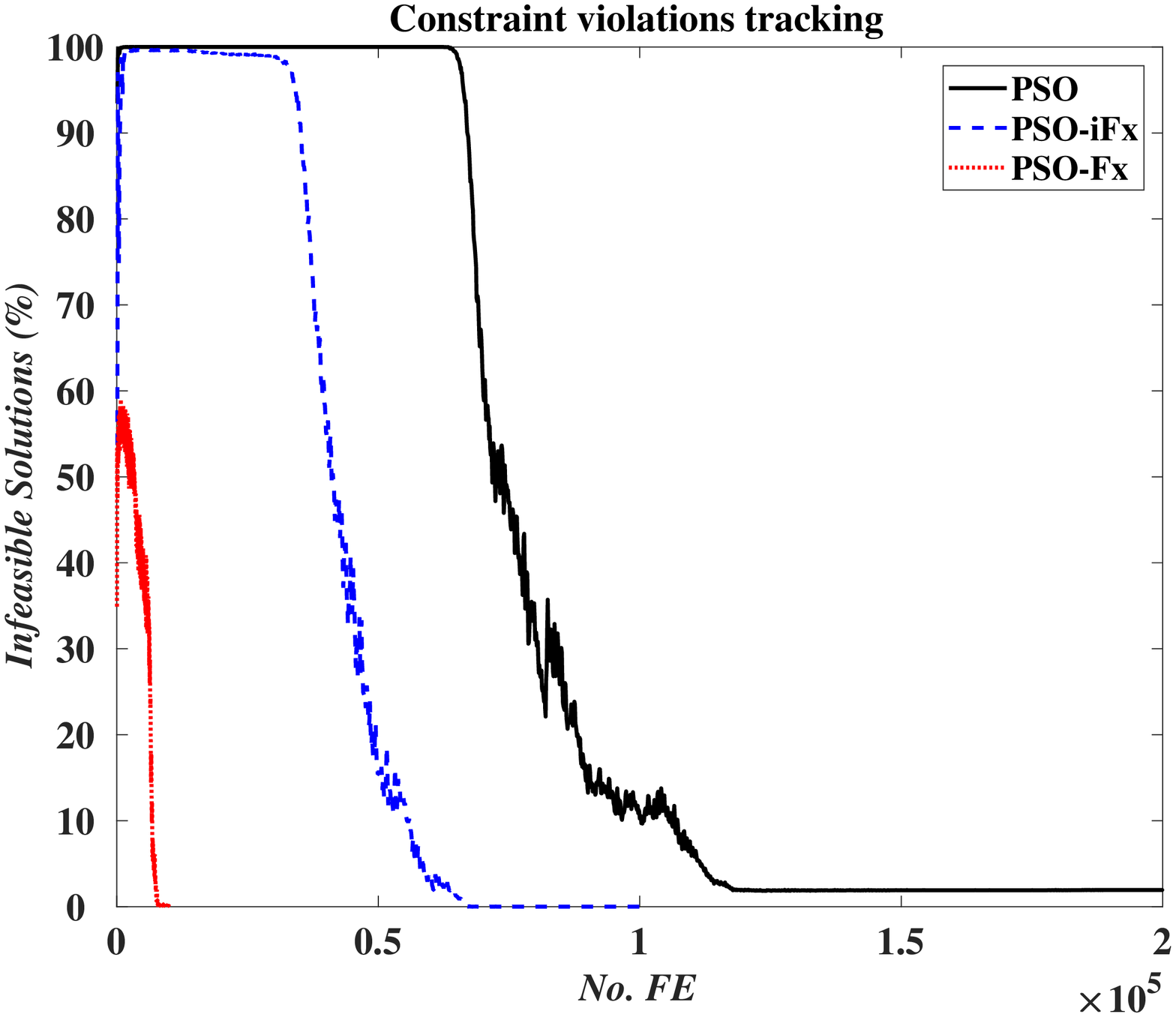,width=4.3cm}\label{sfig:infpso}}\;
          \subfigure[]{\psfig{figure=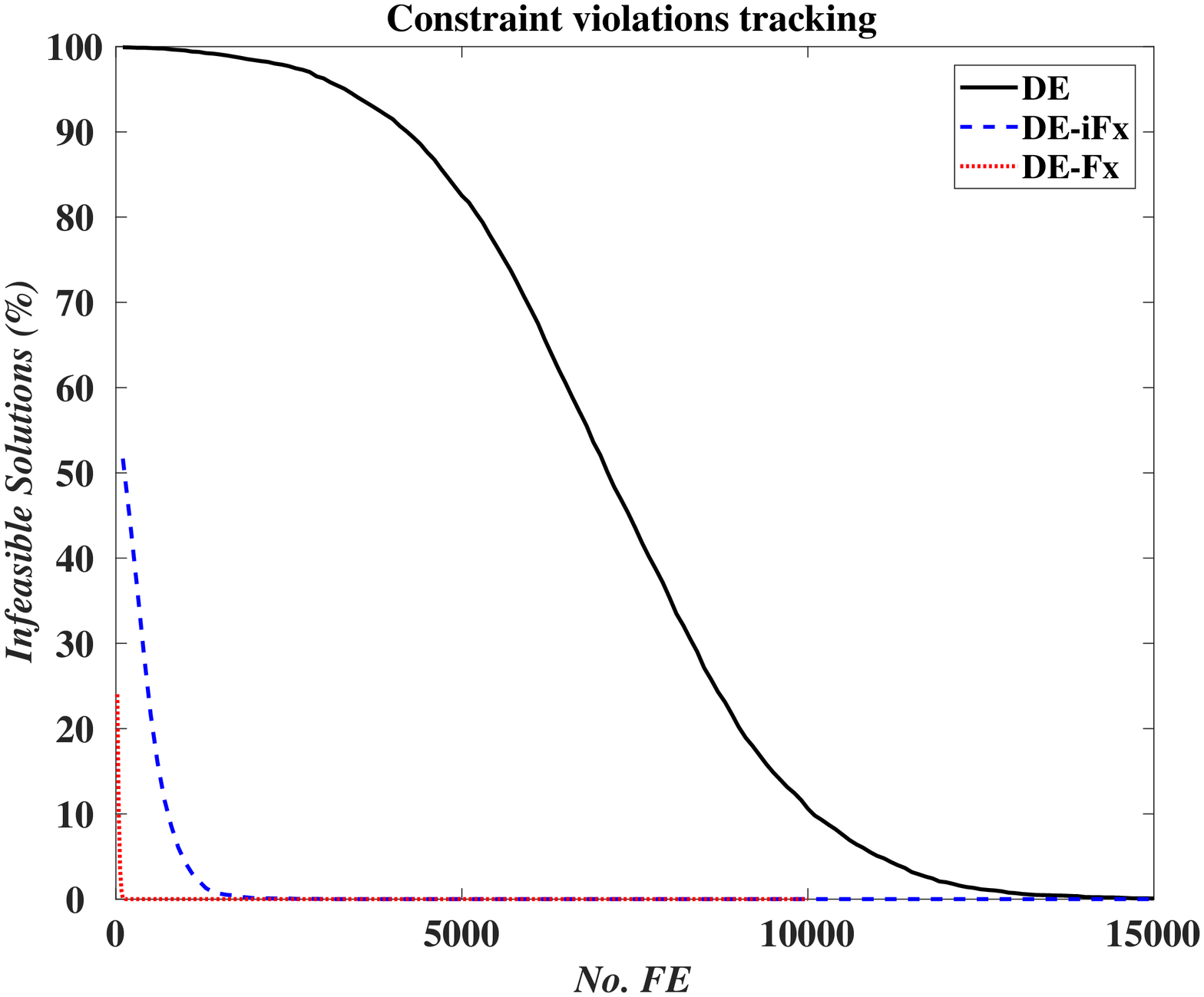,width=4.3cm}\label{sfig:infde}}\;}
\caption{History results (performance plots) of the optimization algorithms for stepped column design problem. a) mean convergence histories of PSO runs, b) mean convergence histories of DE runs, c) histories of infeasible candidate solutions rate for PSO runs, d) histories of infeasible candidate solutions rate for DE runs.}
\label{fig:convplot}
\end{figure}

Using the new formula, the problem with 50 variables is converted to a problem with only two variables. This problem has been solved using both PSO and DE algorithms, and each algorithm has been used for optimizing this problem using three strategies, as follows:
\begin{enumerate}
\item Optimization algorithm for solving the problem without reformulation 
\item Optimization algorithm for solving the problem with reformulation only for initialization ($iFx$)
\item Optimization algorithm for solving the problem with reformulation ($Fx$)
\end{enumerate}
Due to the random nature of the optimization process, each strategy runs 51 times, and the performance plots are presented in Figure~\ref{fig:convplot}.
At first, it is obvious that the PSO and DE algorithms have completely different results in optimizing a constrained problem, which comes from the different nature of the algorithms. Also, it should be noted that, because of the selection operator in DE, a feasible solution cannot be replaced with an infeasible solution. Thus, the number of infeasible solutions cannot increase and the DE histories are smooth.
Conversely, the PSO algorithm does not have such a selection operator, and because of that, a feasible solution can be replaced with an infeasible solution and as a results the number of infeasible solutions may increase during the search process. 

From Figures~\ref{sfig:pso} and~\ref{sfig:de}, it can be seen that $iFx$ and $Fx$ strategies have better starting points.
This should be the main reason why they converge more quickly and to better values, in comparison to the original algorithm without the variable functioning approach.
From these two performance plots, it can be seen that the third strategy can converge after a few iterations.
Since the $Fx$ strategy reduces the problem dimension from 50 to 2, it converges very fast.
Although the third strategy converges very quickly, its final solution is slightly worse than the second strategy for this problem, due to the fact that the candidate solutions being forced to stay with the defined function during all iterations.
From Figures~\ref{sfig:infpso} and~\ref{sfig:infde}, all candidate solutions to the algorithms with the proposed strategy are quickly converging on the feasible region, and their convergence rates are better than those of $iFx$ and $Fx$ strategies.
The PSO-iFx candidate solutions move toward the feasible region faster than those of PSO, which shows initialization using the proposed approach could be effective in PSO.
Comparing the DE and DE-iFx, it can be seen that DE-iFx candidate solutions go toward the feasible region much faster. 
For this stepped column example, it is clear that PSO and DE results are significantly improved after using the proposed approach.
However, the steel frame design optimization is more complex, and cross-section areas should be selected from predefined sections.
In the next section, steel frame design optimization cases are used to investigate the applicability of the proposed method.

\section{Numerical Case Studies}
\label{sec:cases}

\begin{figure*}[t]
\centering
    \mbox{
\subfigure[1-bay 8-story frame ]{\psfig{figure=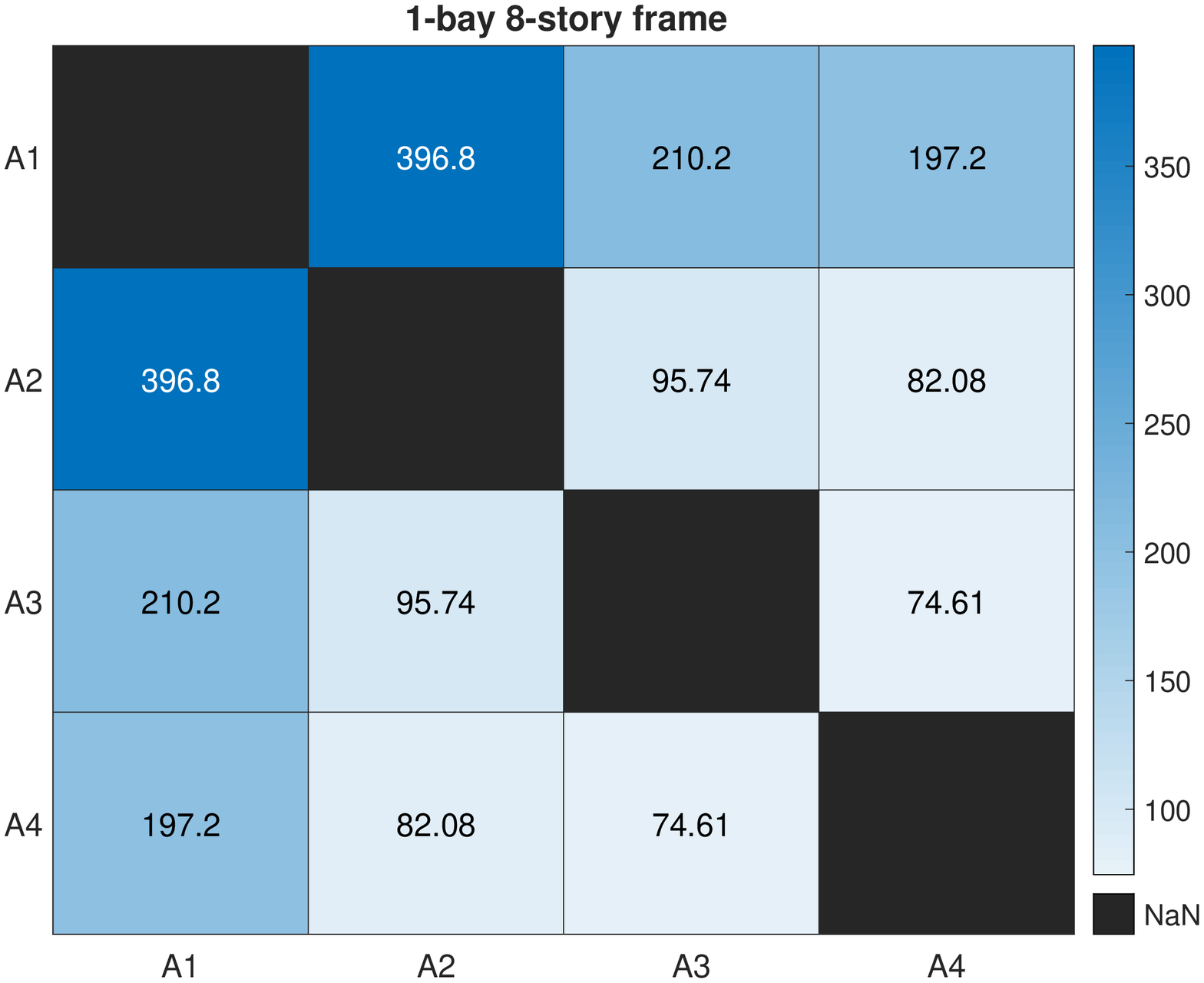 ,width=.3\textwidth}\label{sfig:dg1b8s}}\;
\subfigure[3-bay 15-story frame]{\psfig{figure=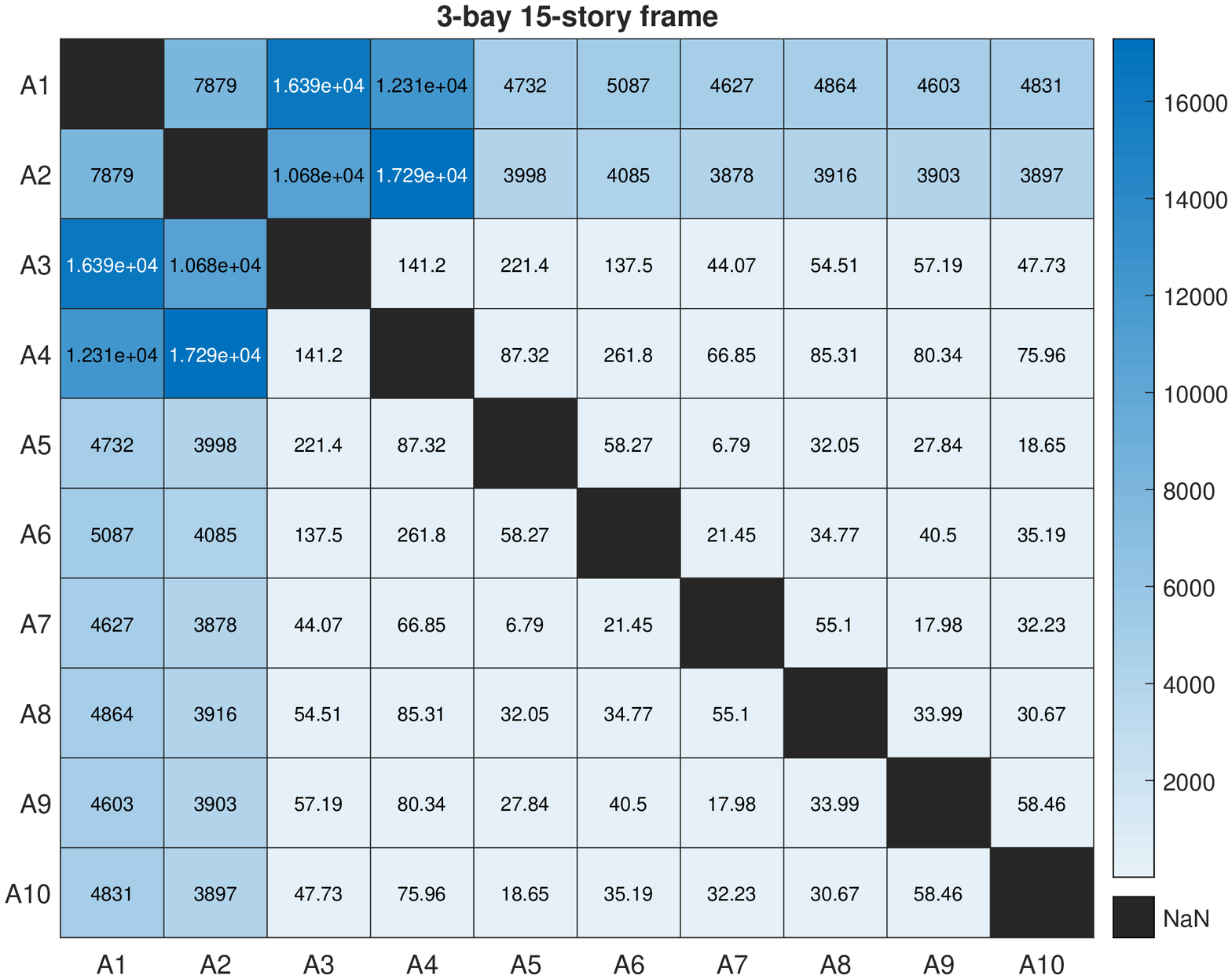,width=.3\textwidth}\label{sfig:dg3b15s}}\;
\subfigure[3-bay 24-story frame]{\psfig{figure=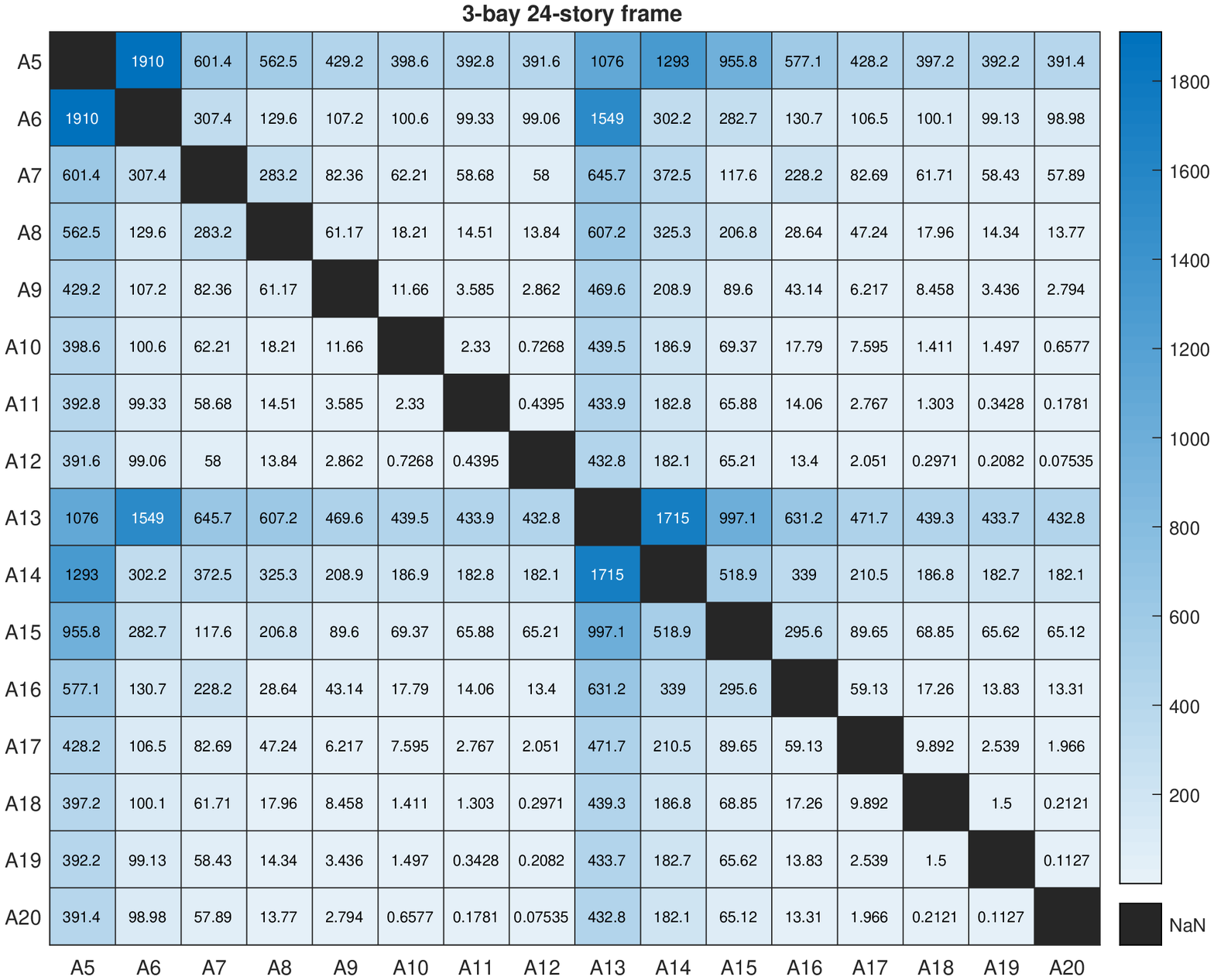,width=.3\textwidth}\label{sfig:dg3b24s}}\;
}
\caption{Variable interaction matrix of various steel frames.}
\label{fig:dgmats}
\end{figure*}

Design of three steel frame structures are optimized in this section as case studies.
In the structural engineering literature, frame structures are usually defined by the number of stories and bays~\cite{talatahari2015optimum}, and therefore these benchmarks for 1-bay 8-story, 3-bay 15-story, and 3-bay 24-story, the best-designed frame case studies, are considered in this study~\cite{talatahari2015optimum}. The level of interaction between each pair of variables for these three case studies are visualized in figure~\ref{fig:dgmats} which confirms that all column cross-section depends on base column cross-section. These steel frame design problems are optimized by both PSO and DE algorithms.
Like the illustrative example, three strategies are implemented in each of the optimization algorithms.
Because of the random nature of the optimization process, each strategy runs 51 times to obtain meaningful results.
After trial and error, with the initial population up to 100, it is found that the best range is 20 to 60 for the benchmark steel frame problems with and without the variable functioning approach, as illustrated in Table~\ref{tab:popsize}.

\begin{table}[t]
\centering
\scriptsize
\caption{Population size of benchmark steel frames.}
\begin{tabular}{lccc}
\toprule
\multirow{2}{*}{Problem} & \multicolumn{3}{c}{Variable Functioning}   \\ \cmidrule{2-4}
                         &  none & $iFx$ &  $Fx$ \\ \midrule
1-bay  8-story frame     & 25            & 25            & 20        \\
3-bay 15-story frame     & 40            & 40            & 25        \\
3-bay 24-story frame     & 60            & 60            & 25        \\
\bottomrule
\end{tabular}
\label{tab:popsize}
\end{table}

The maximum number of function evaluations, in other words, the number of finite element (FE) analyses, is considered to be the stopping criterion, which can be found from the convergence history plots.
The structural analysis section is also coded in MATLAB, using the matrix stiffness method implementation of FE.

\subsection{Design of a 1-bay, 8-story frame}
\begin{figure}[t]
\centering
\includegraphics[width=.5\linewidth]{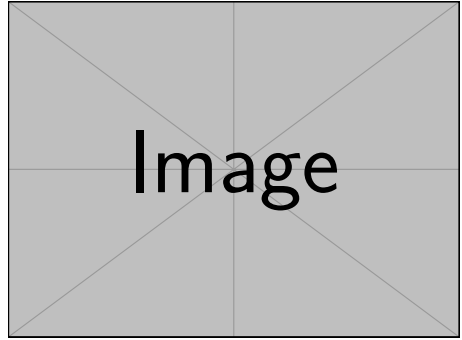}
\caption{Topology of the 1-bay 8-story frame.}
\label{fig:8story}
\end{figure}

The 1-bay 8-story problem is one of the benchmark structural engineering problems~\cite{gandomi2011benchmark}, and has been widely used in the literature (e.g.~\citet{juliani2022efficient}).
The configuration of this frame structure, including the applied loads, is shown in Figure~\ref{fig:8story}. After considering the fabrication conditions affecting the construction of the frame structure, the same beam/column cross-sections are used for the two following stories. The values of both the beam and column element groups are chosen from all 267 W-shapes. In this case, the roof drift is the only performance constraint which should be less than 5.08 cm. The modulus of elasticity (E) of the steel is taken as 200 GPa.

\begin{figure}[t]
\centering
\mbox{
\subfigure[Mean of PSO runs]{\psfig{figure=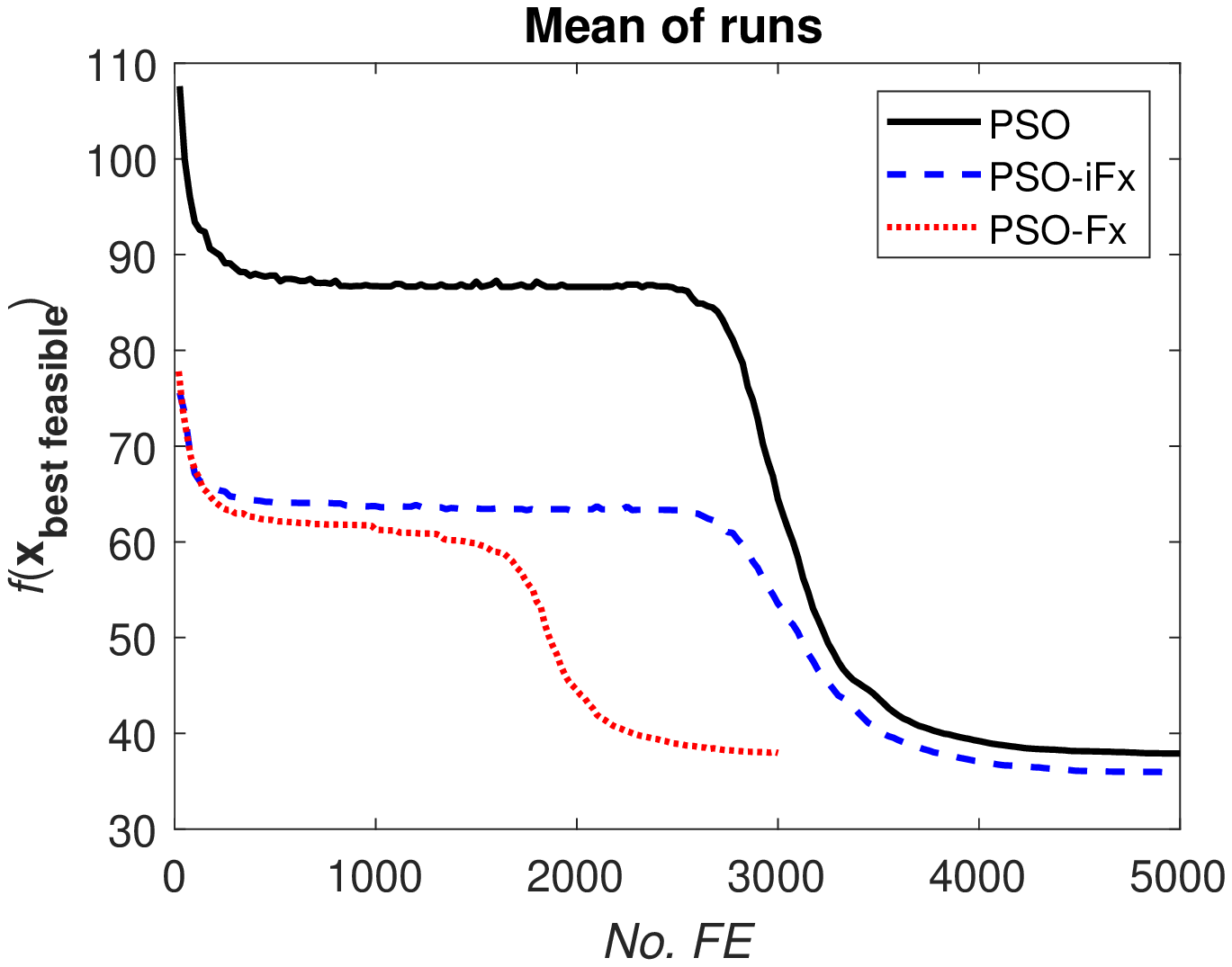,width=4.3cm}\label{sfig:mpso8story}}\;
\subfigure[Mean of DE runs ]{\psfig{figure=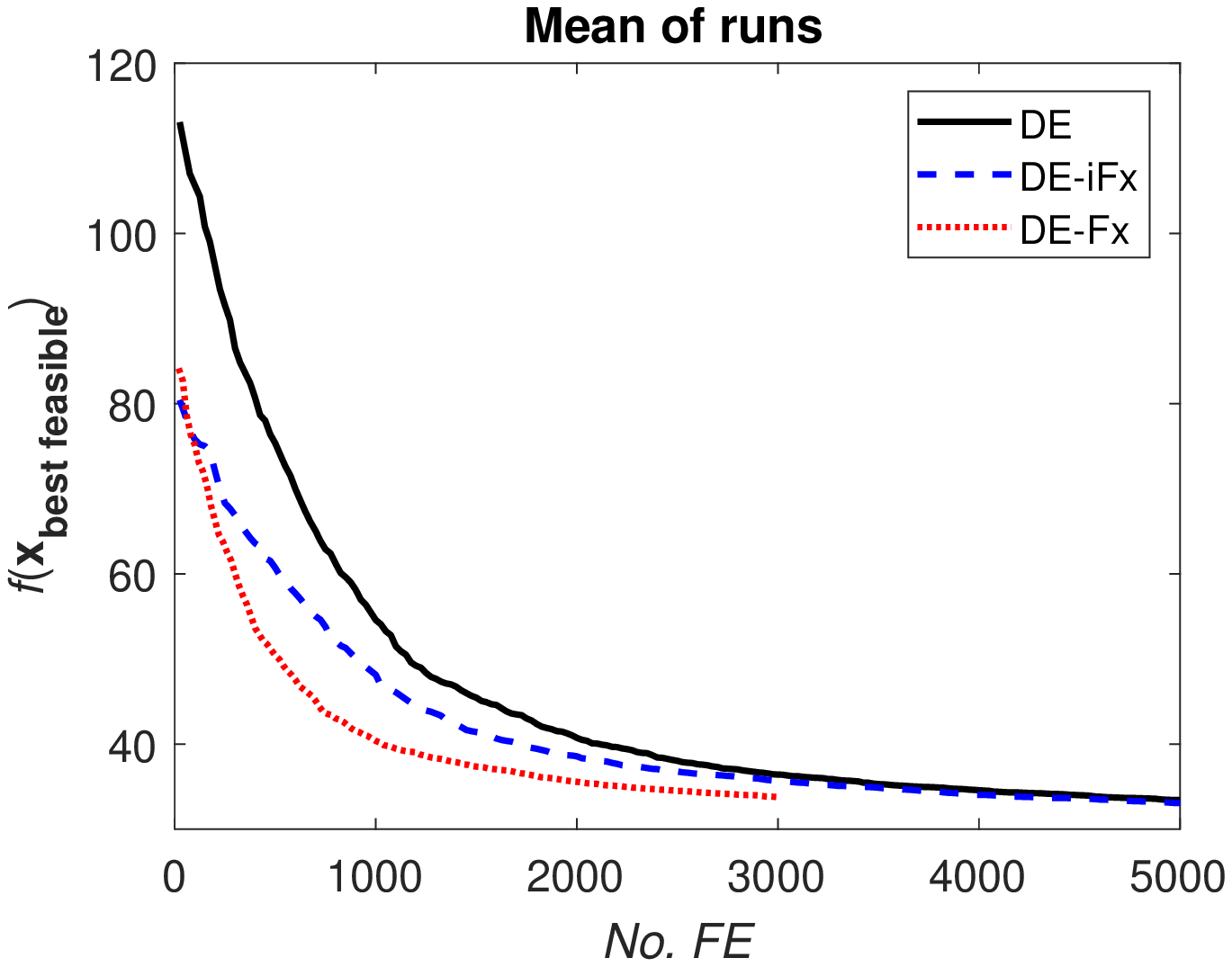,width=4.3cm}\label{sfig:mde8story}}\;
}
\caption{Convergence histories of the 1-bay 8-story frame optimization problem using PSO and DE algorithms.}
\label{fig:conv8story}
\end{figure}

The proposed approach had been used for this problem to reformulate the problem variables (column cross-section areas).
Here, four column cross-section areas are replaced with the base cross-section area and $\alpha$; therefore, the number of column variables is decreased to two variables.
PSO and DE algorithms have solved this benchmark problem with the three defined strategies.
The convergence histories of PSO and DE algorithms with 3000 FEs for the Fx and 5000 FEs for the other strategies and the mean of 51 runs are presented in Figure~\ref{fig:conv8story}.
From the shown histories, it can be seen that the algorithms using $iFx$ and $Fx$ strategies start the search from a better fitness, compared to the algorithm without variable functioning.
These clearly show that the initialization itself ($iFx$) can significantly improve the search process.
Note that each of these methods converges to a different solution and the final cross-sections of any two strategies are not the same.
From PSO convergence history (Figure~\ref{sfig:mpso8story}), it is clear that not only does $iFx$ help the algorithm to converge more quickly, but it also converges to better solutions.
Using the proposed approach during the search process ($Fx$) significantly improves the convergence rate of PSO in comparison with the other two strategies.
The DE convergence histories also have the same trend as those of PSO histories.
Although the convergence rates are improved after using DE-$iFx$ and DE-$Fx$, the improvements are not as significant as those of PSO.
Note that the results of the PSO-$Fx$ and DE-$Fx$ algorithms are obtained after 3,000 FEs, and the others were received after 5,000 FEs.
Therefore, the third strategy ($Fx$) improves the convergence rates  significantly for this case. 

\begin{figure}[t]
    \centering
     \mbox{\subfigure[PSO]{\psfig{figure=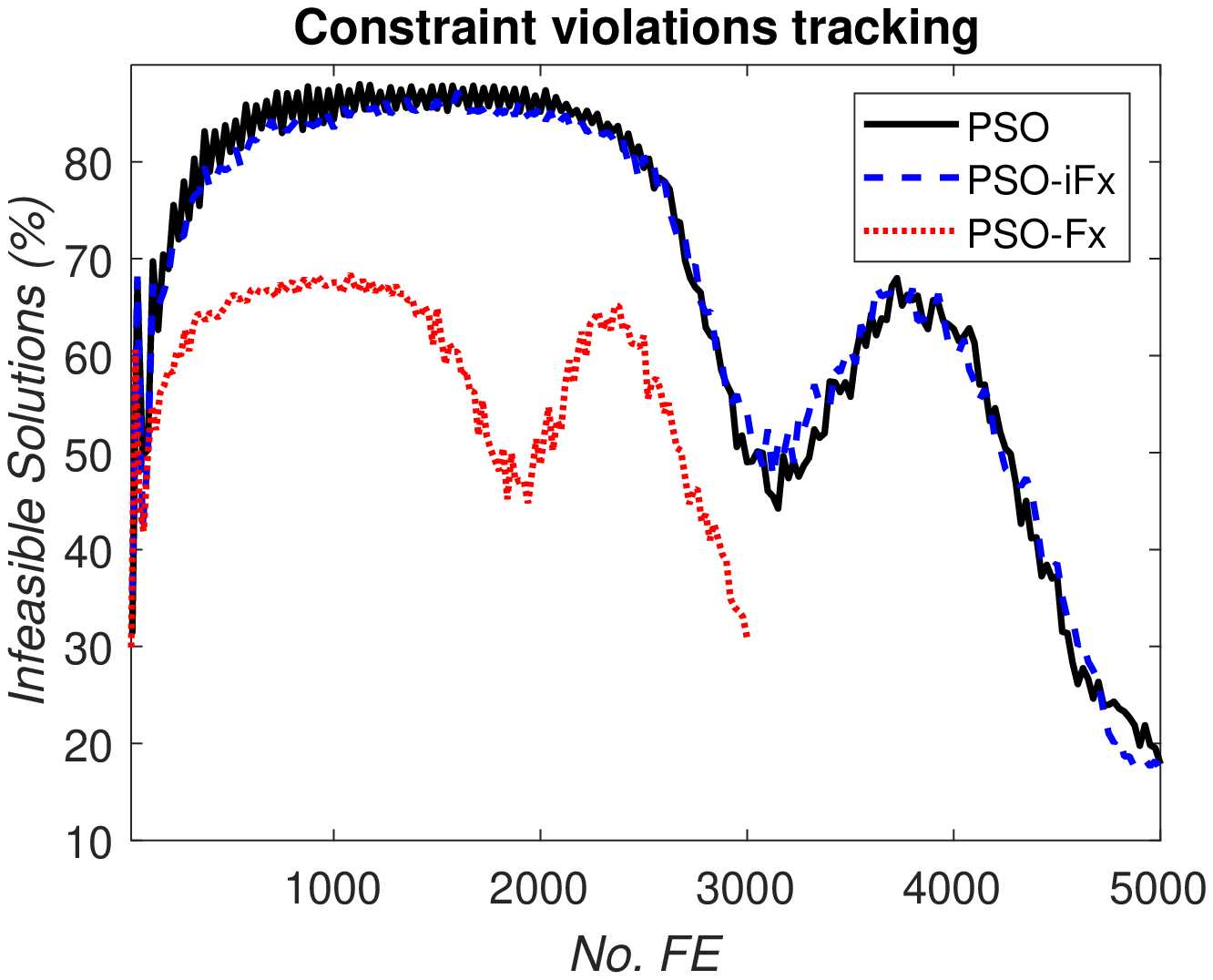,width=4.3cm}\label{sfig:psoinf8story}}\;
           \subfigure[DE ]{\psfig{figure=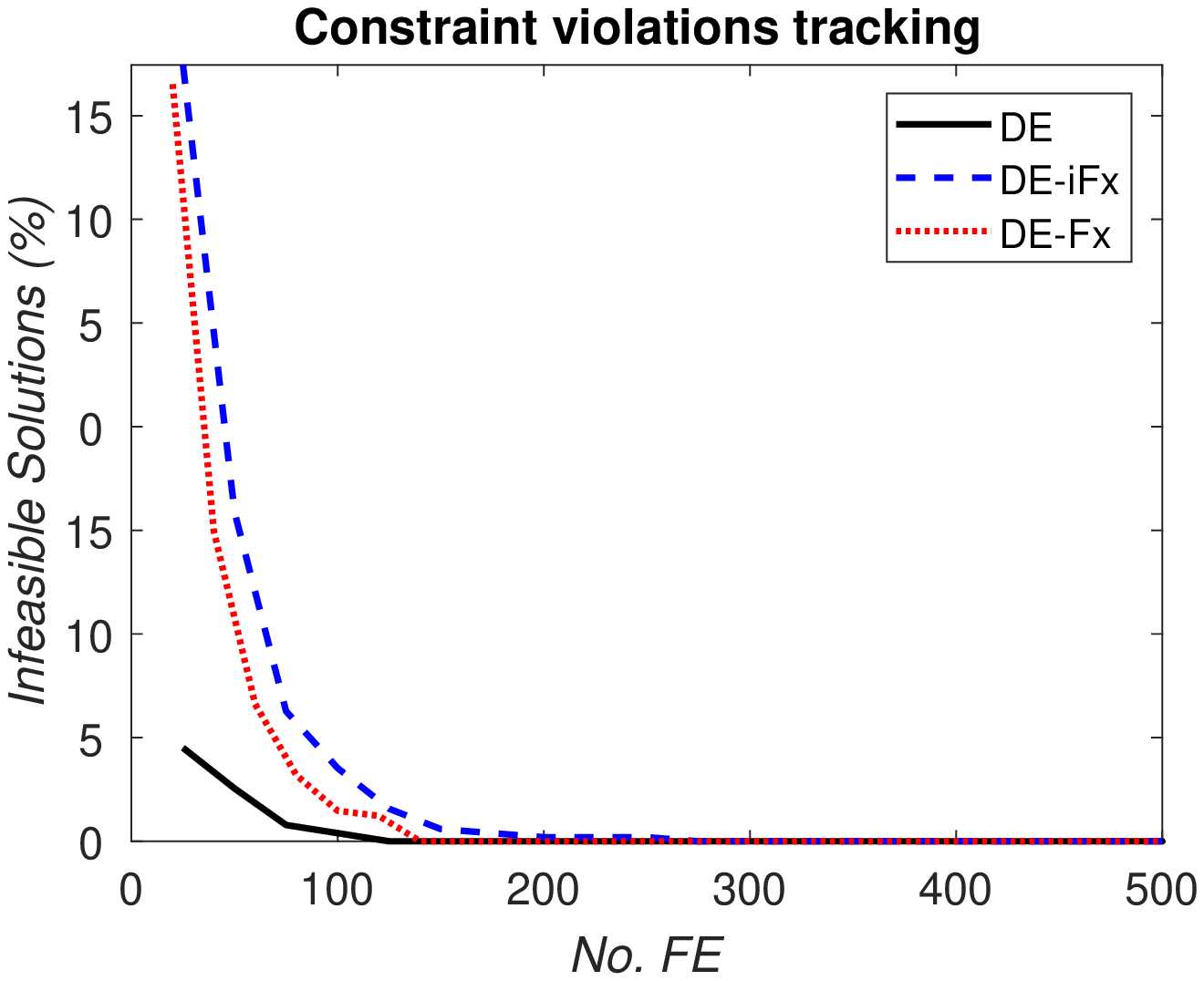,width=4.3cm}\label{sfig:deinf8story}}\;}
\caption{Infeasible solutions histories in 1-bay 8-story frame optimization problem.}
\label{fig:inf8story}
\end{figure}

Figure~\ref{fig:inf8story} tracks the number of infeasible solutions during the PSO and DE iterations, respectively.
From Figure~\ref{sfig:psoinf8story} and Figures~\ref{sfig:mpso8story}, it can be seen that PSO mostly does exploration for about first half of iterations since the best feasible solutions and violation rates almost remains constant after the a few iterations.
Figure~\ref{sfig:psoinf8story} also shows that the percentage of infeasible solutions in PSO and PSO-$iFx$ are similar. This means despite different initializations, both algorithms have the same constraint violation histories.
In Figure~\ref{sfig:psoinf8story}, the PSO-$Fx$ results have the same pattern as the other methods with smaller scale (fewer FEs), which confirms the advantages of $Fx$ strategy over the other two strategies in reaching more feasible solutions.
Figure~\ref{sfig:deinf8story} has an entirely different paradigm, where the constraint violation of DE is less than DE-$iFx$ and DE-$Fx$.
Also, DE-$iFx$ and DE-$Fx$ infeasible solutions histories are very similar, and they both are different from the DE history.
This is because the DE-$iFx$ and DE-$Fx$ strategies generate more infeasible solutions in early iterations in comparison with DE strategy (approximately 10\%).

\subsection{Design of a 3-bay, 15-story frame}

\begin{figure}[t]
\centering
\includegraphics[width=.5\linewidth]{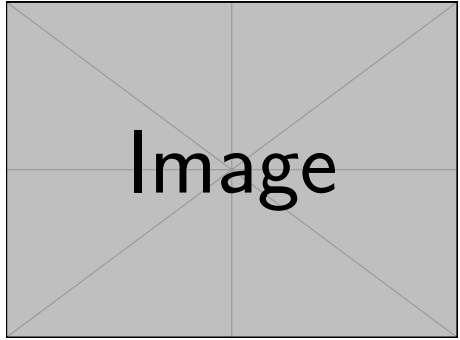}
\caption{Topology of the 3-bay 15-story frame.}
\label{fig:3b15s}
\end{figure}

The 3-bay 15-story frame structure is shown in Figure~\ref{fig:3b15s}, including configuration and applied loads. This steel frame optimization problem have been already used by many researchers as a benchmark~\cite{mosharmovahhed2021design}.
The AISC combined strength and displacement limit (sway of the roof is restricted to 23.5 cm) is considered the constraint for optimizing the frame weight. The used steel has E = 200 GPa and a yield stress (Fy) of 248.2 MPa. After considering the fabrication conditions to the construction of the frame, the same column cross-sections are used for the three following stories.
Because of the symmetry of the frame structure, two sets of columns are considered here: inner columns and outer columns.
Each column set is replaced with one function in order to apply the proposed approach and, as a result, the number of variables is decreased from 11 to 5 after using the $Fx$ approach. 

\begin{figure}[t]
\centering
\mbox{
\subfigure[Mean of PSO runs]{\psfig{figure=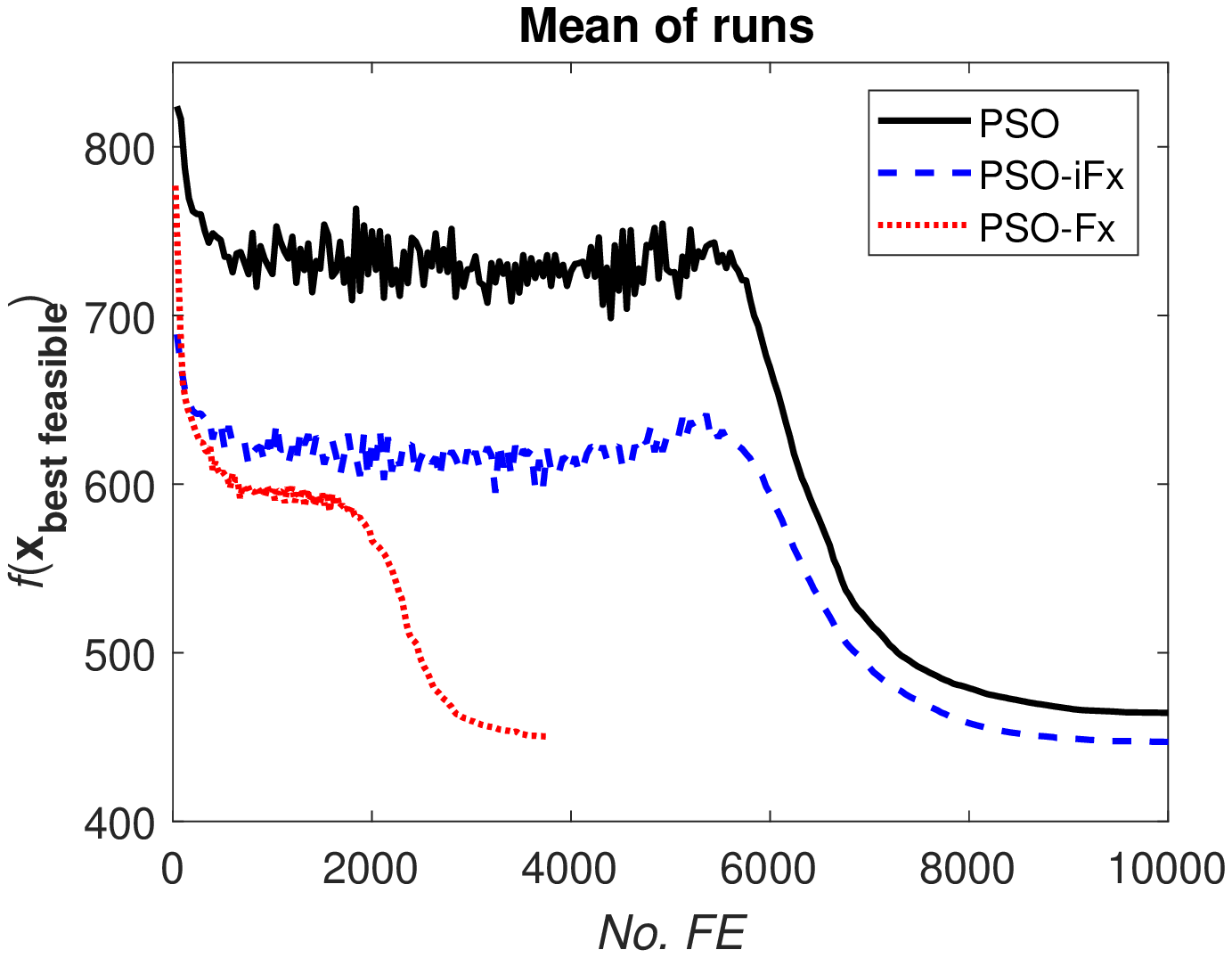,width=4.3cm}\label{sfig:mpso3b15s}}\;
\subfigure[Mean of DE runs ]{\psfig{figure=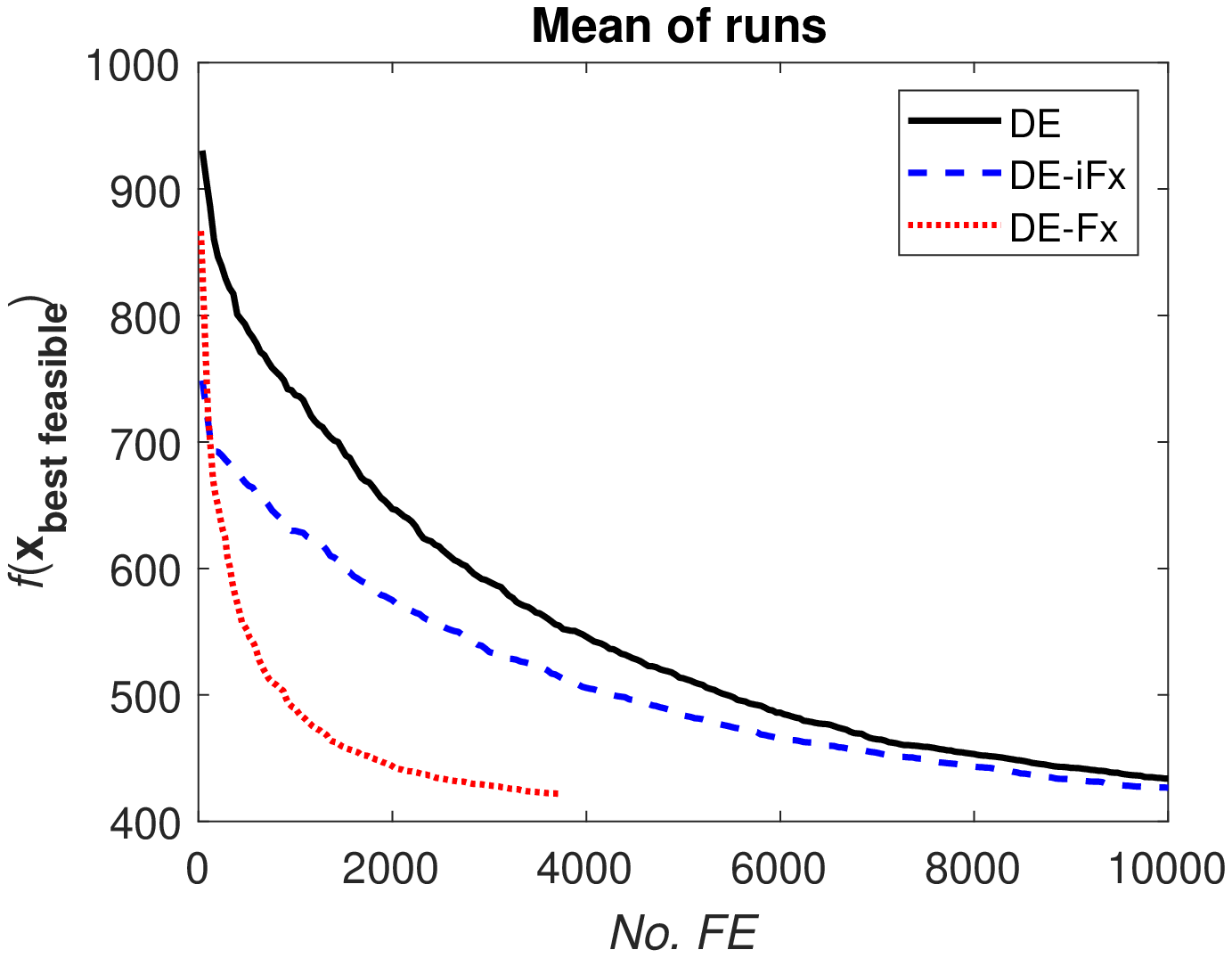,width=4.3cm}\label{sfig:mde3b15s}}\;
}
\caption{Convergence histories of the 3-bay 15-story frame optimization problem using PSO and DE algorithms.}
\label{fig:conv3b15s}
\end{figure}

The 3-bay 15-story frame problem has been solved by PSO and DE algorithms with different strategies.
The maximum number of FEs is set to 4,000 for the Fx strategy and 10,000 is set for the maximum number of FEs of the other strategies. Also, it should be clarified that each of these methods converges to a different solution which is expected due to the vast search space of this case study.
The convergence histories of PSO and DE algorithms for the mean of runs are presented in Figure~\ref{fig:conv3b15s}.
From Figure~\ref{sfig:mpso3b15s}, it can be seen that PSO-$iFx$ convergence is slightly improved, in contrast with PSO.
However, it is clear that PSO-$iFx$ notably improves the average results (shown in Figure~\ref{sfig:mpso3b15s}).
Using the proposed approach during the whole iterations ($Fx$), significantly improves the convergence of PSO for this problem in comparison with the other two strategies.
Similar to the PSO results, it can be seen that the $iFx$ strategy improves the DE algorithm, and using the approach during all iterations ($Fx$) significantly improves the convergence history (Figure~\ref{sfig:mde3b15s}).
Therefore, it is clear that using the proposed approach can notably improve the results for this steel frame optimization problem. 

From Figure~\ref{sfig:mpso3b15s}, it can be said that PSO-$Fx$ and PSO-$iFx$ methods need fewer than 3,000 and 8,000 FEs, respectively, to reach the objective value obtained by PSO after 10,000 FEs.
As it is shown in Figure~\ref{sfig:mde3b15s}, the same thing happened for the DE based algorithms, as both DE-$Fx$ and DE-$iFx$ methods reached better objective values in a lesser FEs.
From Figure~\ref{sfig:mde3b15s}, the improvement is significant for DE-$iFx$, which only needs about a quarter of the FEs in comparison with DE methods to reach the same objective value. 

\begin{figure}[t]
    \centering
     \mbox{\subfigure[PSO]{\psfig{figure=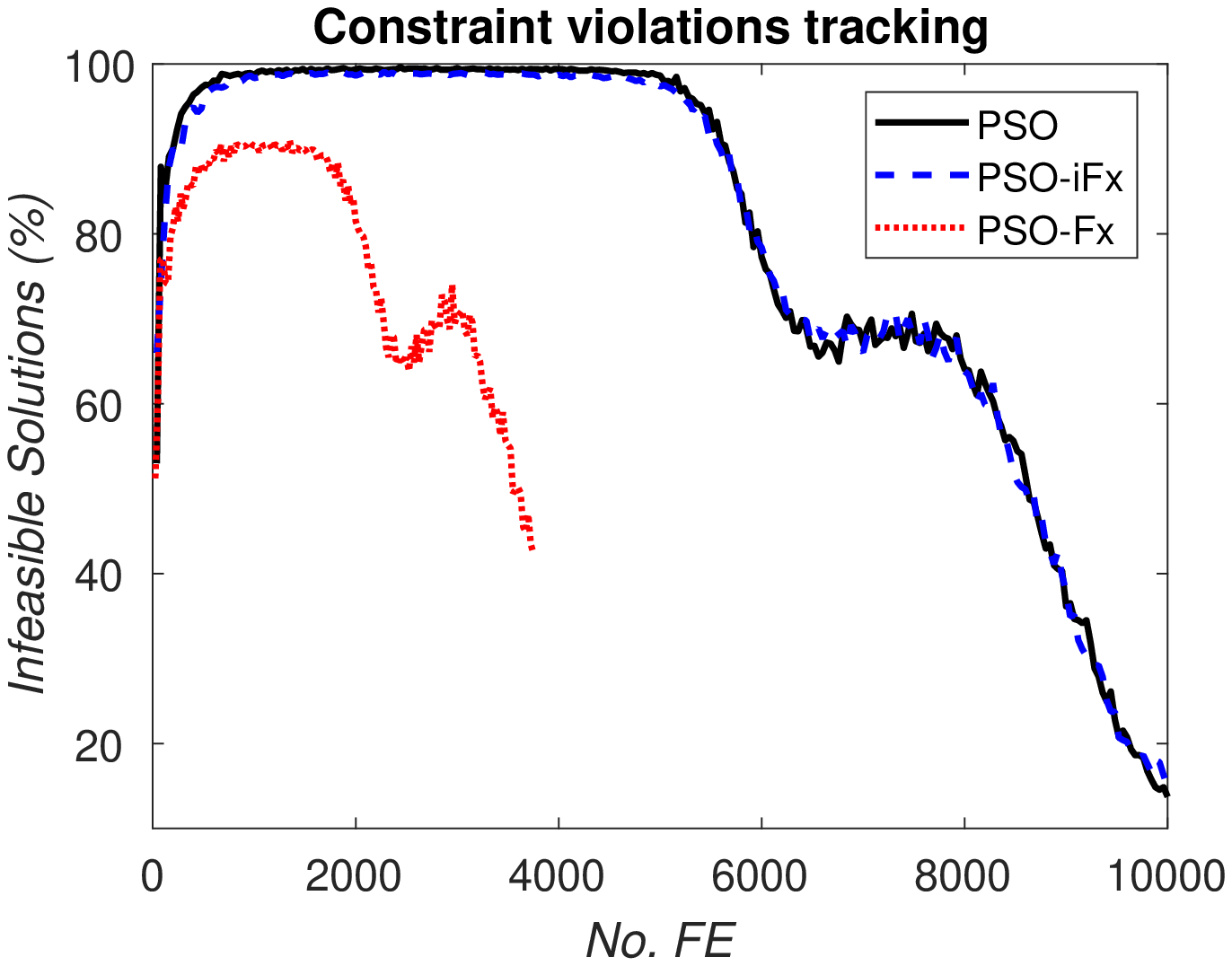,width=4.3cm}\label{sfig:infpso3b15s}}\;
           \subfigure[DE ]{\psfig{figure=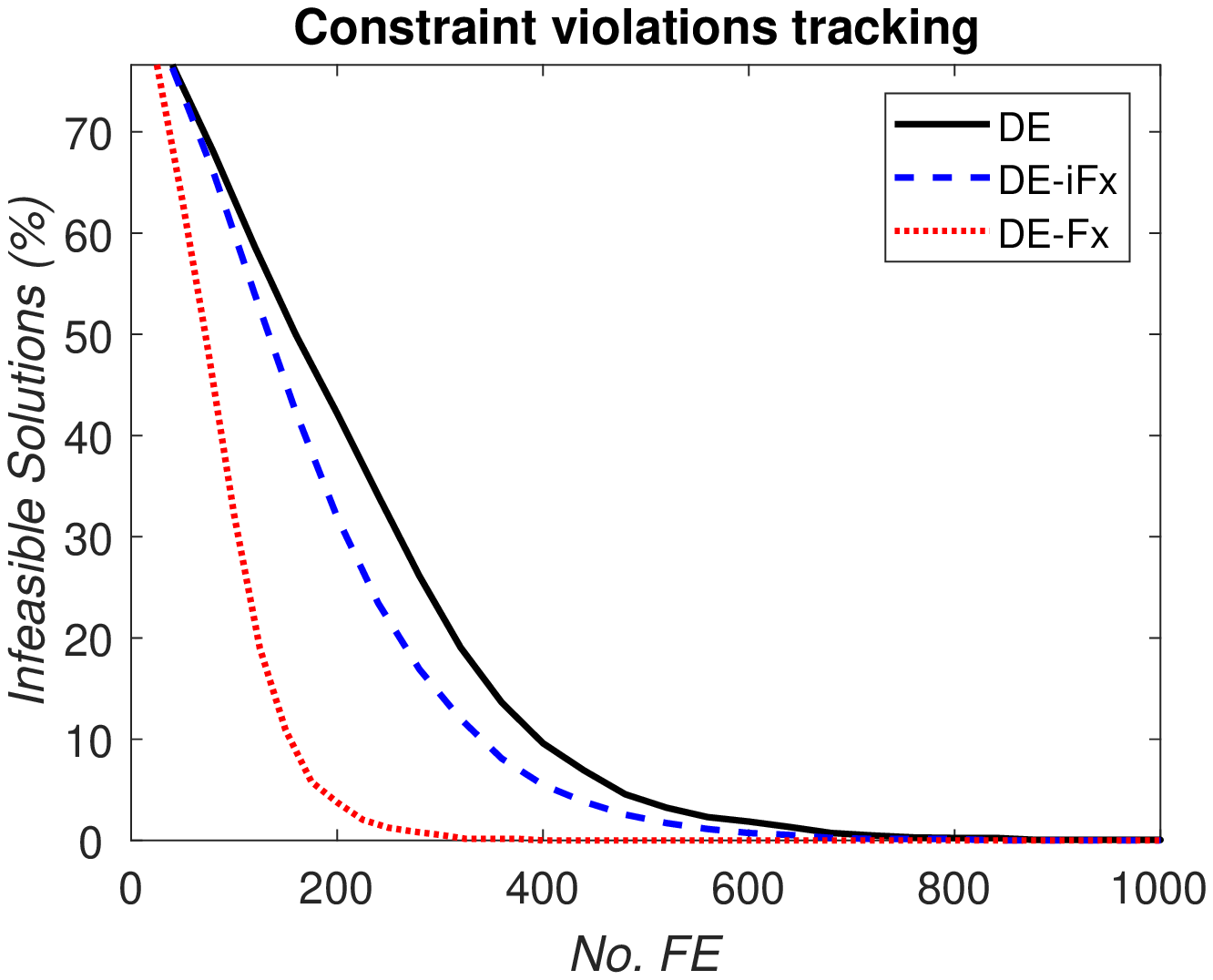,width=4.3cm}\label{sfig:infde3b15s}}\;}
\caption{Infeasible solutions histories in 3-bay 15-story frame optimization problem.}
\label{fig:inf3b15s}
\end{figure}

Figure~\ref{fig:inf3b15s} tracks the number of infeasible solutions during iterations in PSO and DE algorithms, respectively.
From Figure~\ref{sfig:infpso3b15s}, it can be seen that the percentage of infeasible solutions of PSO and PSO-$iFx$ during optimization processes have a similar pattern.
This shows that despite different initializations, these two algorithms have the same constraint violation histories.
In Figure~\ref{sfig:infpso3b15s}, the PSO-$Fx$ results have the same pattern as the other methods with fewer FEs. This confirms the advantages of the proposed approach to finding more feasible solutions when it is used during iterations.
However, using the proposed method for the initialization (shown in Figure~\ref{sfig:infde3b15s}) slightly increases the number of feasible solutions, in comparison with the DE strategy.
Once again, using the proposed approach during iterations significantly enhance the convergence to the feasible region.

It can be seen that using the proposed method for either the initialization or whole iterations result in cost savings. For PSO algorithm, using $iFx$ strategy leads to the best results, with 3.6\%  improvement of median over the 51 PSO runs. However, using $Fx$ during whole search  process only makes 2.5\% improvement of median. The DE results show that $Fx$ strategy could produce the most improvement (3.1\%) of the median and it is interesting since it only uses 3,750 FEs, which is far fewer than other strategies, which use 10,000 FEs. It should be noted that if we only use the function for seeding of DE algorithm, it still has about 2\% improvement of the median.


\subsection{Design of a 3-bay 24-story frame}

\begin{figure}[t]
\centering
\includegraphics[width=.5\linewidth]{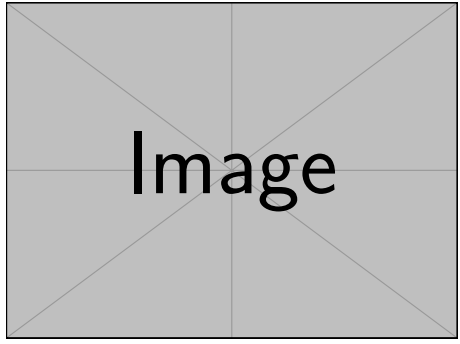}
\caption{Topology of the 3-bay 24-story frame.}
\label{fig:3b24s}
\end{figure}

The topology of the 3-bay 24-story frame, along with the service loading conditions, are shown in Figure~\ref{fig:3b24s}. This tall steel frame problem originally designed by~\citet{davison1974stability} and has 168 members. This case study has later been optimized by many researcher as a challenging steel frame optimization problem~\cite{bigham2020topology}. After imposing the fabrication conditions on the construction of the frame, the same beam cross-section is used in the first and third bay of all floors, except the roof beams.
The interior columns are combined into one set, and the exterior columns are combined in another set of three following stories, which results in 20 design variables -- four beam and 16 column groups.
This steel frame structure is designed based on the LRFD specifications, under inter-story drift constraints.
The steel material has $E = 205$ GPa and $Fy = 230.3$ MPa.
In this problem, column element groups should be chosen from W14 sections (37 W-shapes), while beam element groups could be any of the 267 W-shapes.

\begin{figure}[t]
    \centering
\mbox{
\subfigure[Mean of PSO runs]{\psfig{figure=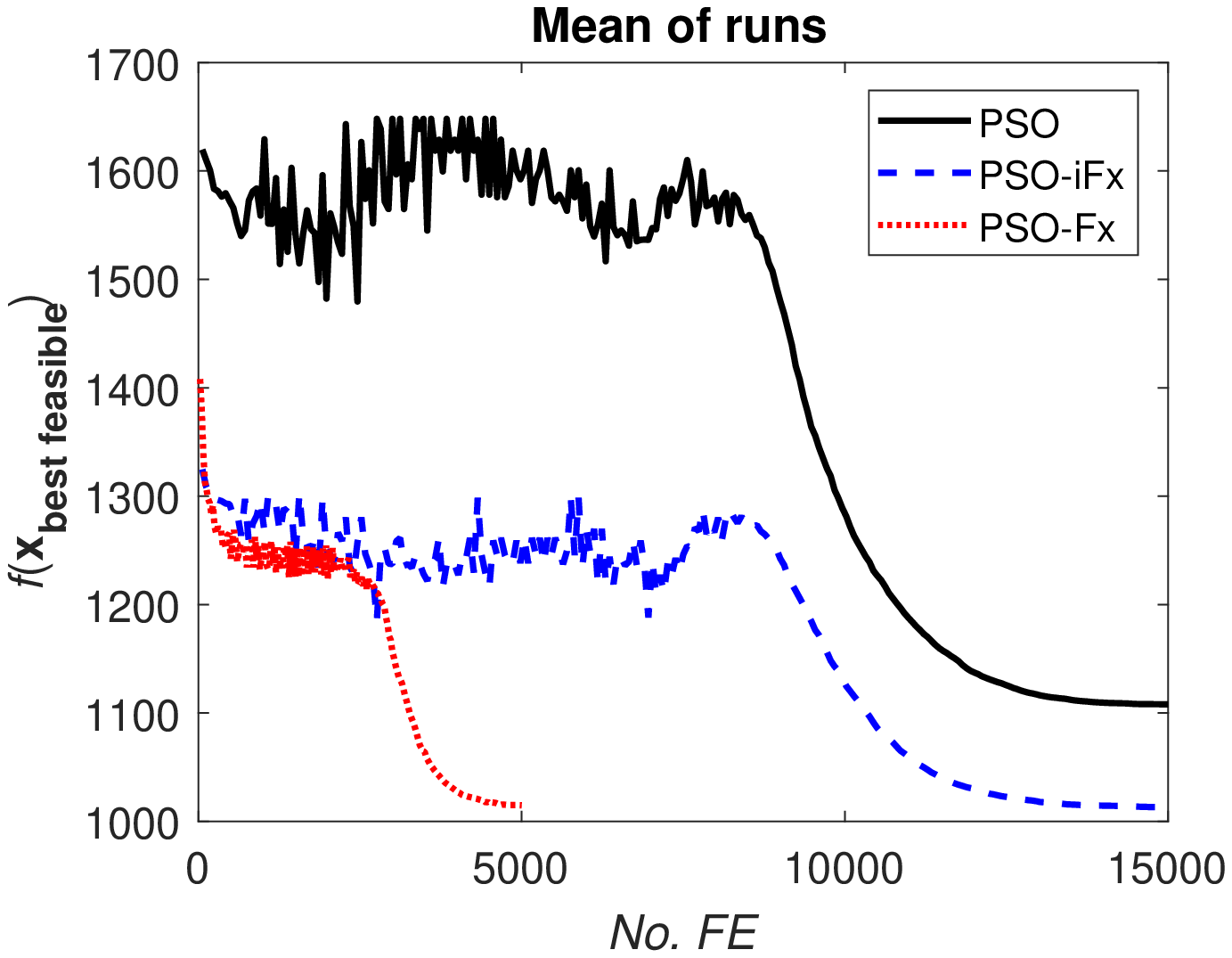,width=4.3cm}\label{sfig:mpso3b24s}}\;
\subfigure[Mean of DE runs ]{\psfig{figure=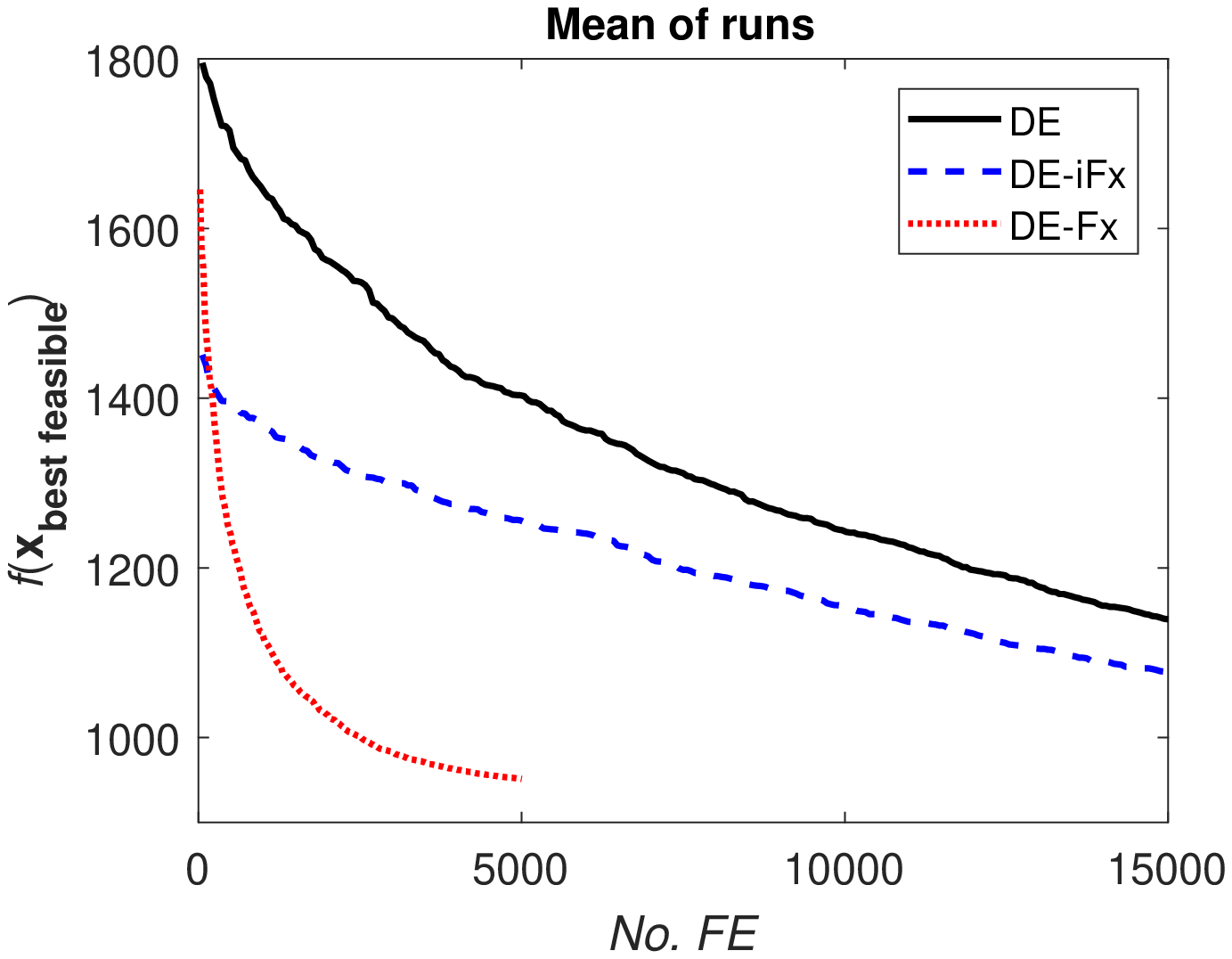,width=4.3cm}\label{sfig:mde3b24s}}\;
}
\caption{Convergence histories of the 3-bay 24-story frame optimization problem using PSO and DE algorithms.}
\label{fig:conv3b24s}
\end{figure}

One function is assigned to each of the column sets, interior column and exterior column sets, and, therefore, the number of variables for the columns are decreased from 16 to 4, which is significant dimension reduction.
The 3-bay 24-story steel frame design problem has been solved by PSO and DE algorithms, with the three defined strategies with the maximum number of iterations of 5,000 and 15,000 for the Fx and other strategies, respectively. 
The convergence histories of PSO and DE algorithms for the mean of runs are respectively presented in Figure~\ref{fig:conv3b24s}.
From these histories, it is clear that both $iFx$ and $Fx$ strategies not only improve the convergence rate of PSO and DE algorithms significantly, but they also converge to better solutions.
Comparing the strategies using the proposed approach, $Fx$ has more improvements in the convergence rate in comparison with $iFx$. 
Each of these methods converges to a different solution which is expected because of the huge search space of the case study.

Figure~\ref{sfig:mpso3b24s} illustrates that the PSO-$Fx$ and PSO-$iFx$ methods need less than 4,000 and 10,000 FEs respectively to reach the objective value of PSO after 15,000 FEs.
As it is shown in Figure~\ref{sfig:mde3b24s}, the same thing happened for DE algorithms, and both DE-$Fx$ and DE-$iFx$ methods reached better objective values using fewer FEs.
From Figure~\ref{sfig:mde3b24s}, the improvement is significant for DE-$iFx$ since it only needs about a quarter of the FEs in comparison with DE methods to reach the same objective value. 

\begin{figure}[t]
    \centering
     \mbox{\subfigure[PSO]{\psfig{figure=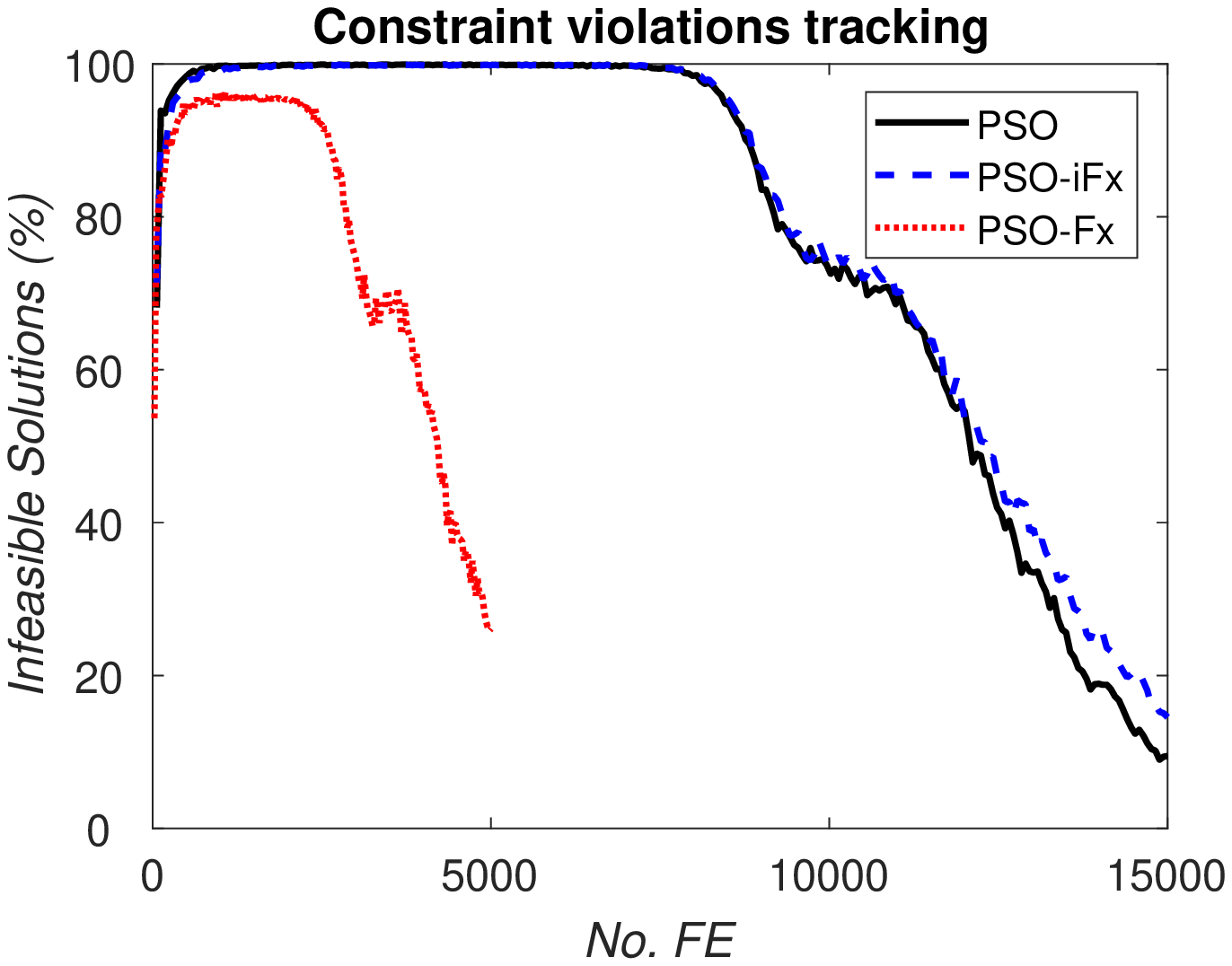,width=4.3cm}\label{sfig:infpso3b24s}}\;
           \subfigure[DE ]{\psfig{figure=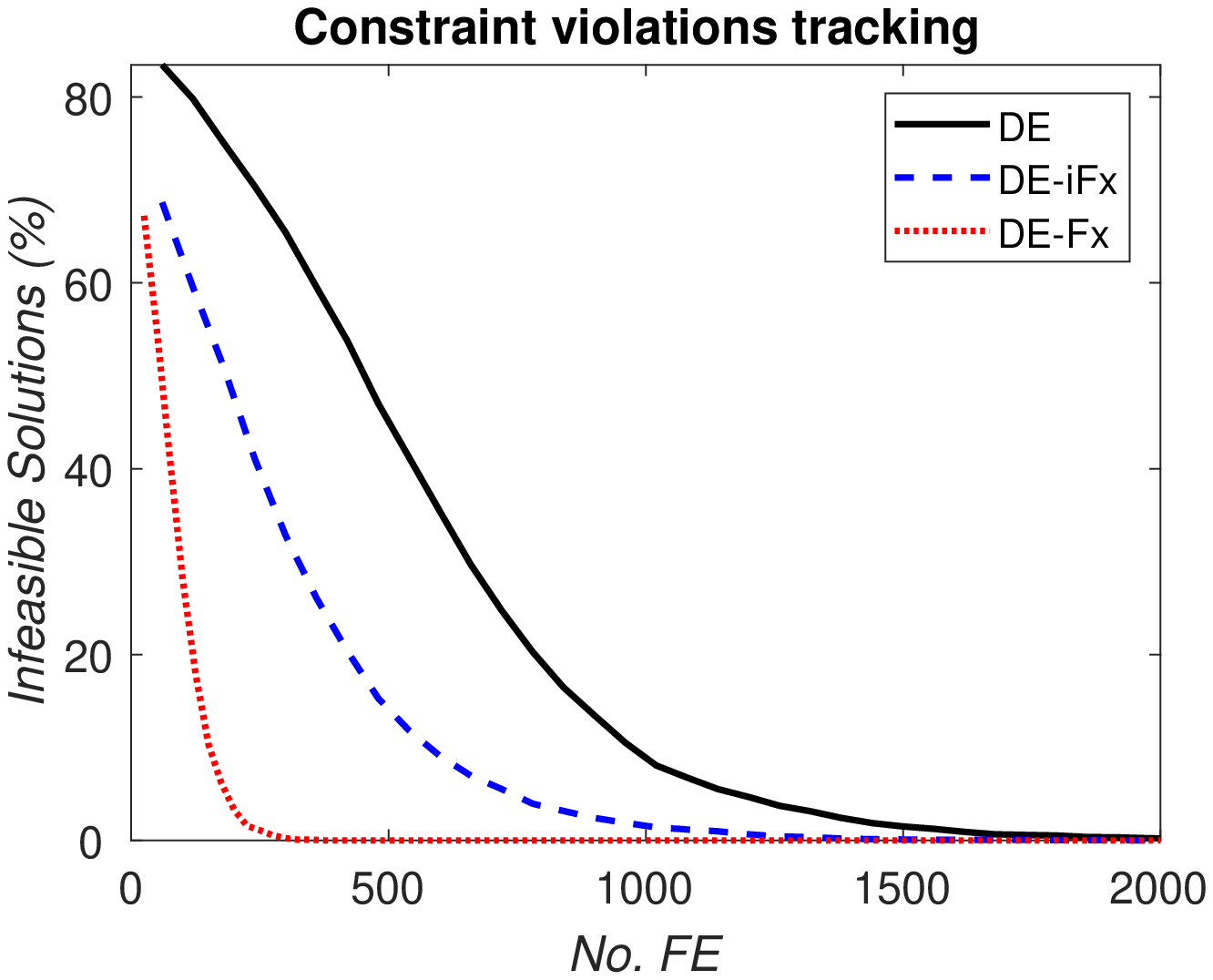,width=4.3cm}\label{sfig:infde3b24s}}\;}
\caption{Infeasible solutions histories in 3-bay 23-story frame optimization problem.}
\label{fig:inf3b24s}
\end{figure}

Figure~\ref{fig:inf3b24s} tracks the percentage of infeasible solutions during the histories in PSO and DE algorithms, respectively.
From Figure~\ref{sfig:infpso3b24s}, it can be seen that the percentages of infeasible solutions of PSO and PSO-$iFx$ during iterations are almost identical.
In Figure~\ref{sfig:infpso3b24s}, the solutions of the PSO-$Fx$ algorithm can move toward the feasible area in fewer FEs, which confirms the advantages of the proposed approach to finding more feasible solutions when it is used during the entire search process.
In Figure~\ref{sfig:infde3b24s}, however, using the proposed approach for the initialization notably increases the number of feasible solutions ($iFx$) and using it during whole iterations ($Fx$) can even improve the number of feasible solutions, and all solutions become feasible at early iterations.

It can be seen that $iFx$ and $Fx$ strategies results in better designs. For PSO algorithm, using both $iFx$ and $Fx$ strategies improve the median results about 9\%. For DE algorithm, with only using the function for seeding of DE algorithm ($iFx$), it  has about 5.9\% improvement of the median. Moreover, $Fx$ strategy is much more effective and improves the median results of DE by 16.8\% with only one third of FEs used in other strategies.

\begin{figure}[t]
    \centering
     \mbox{\subfigure[Outer column using PSO]{\psfig{figure=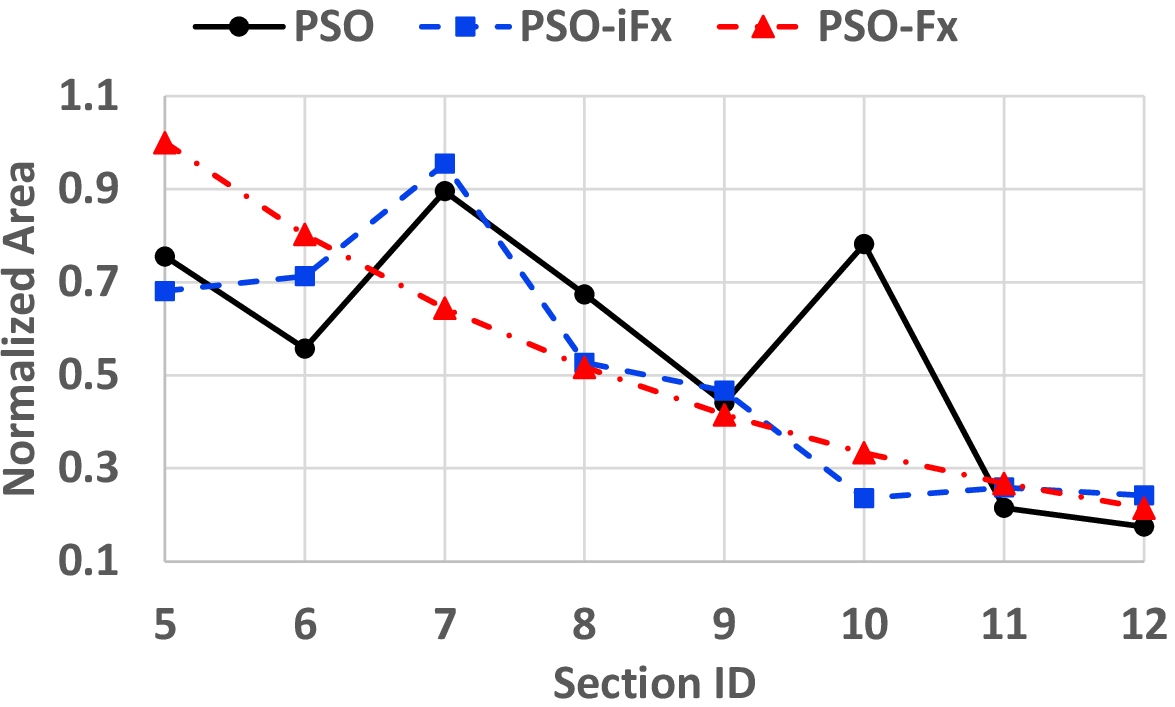,width=4.3cm}\label{sfig:outpso3b24s}}\;
           \subfigure[Inner column using PSO]{\psfig{figure=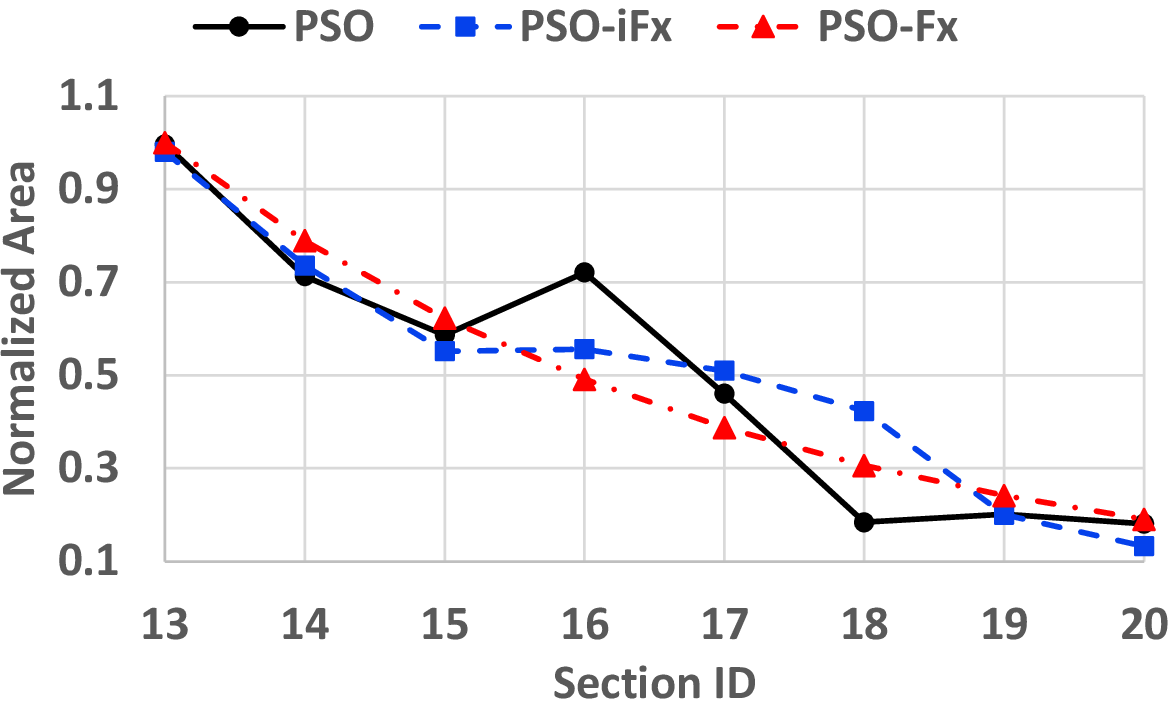,width=4.3cm}\label{sfig:inpso3b24s}}\;}
     \mbox{\subfigure[Outer column using DE ]{\psfig{figure=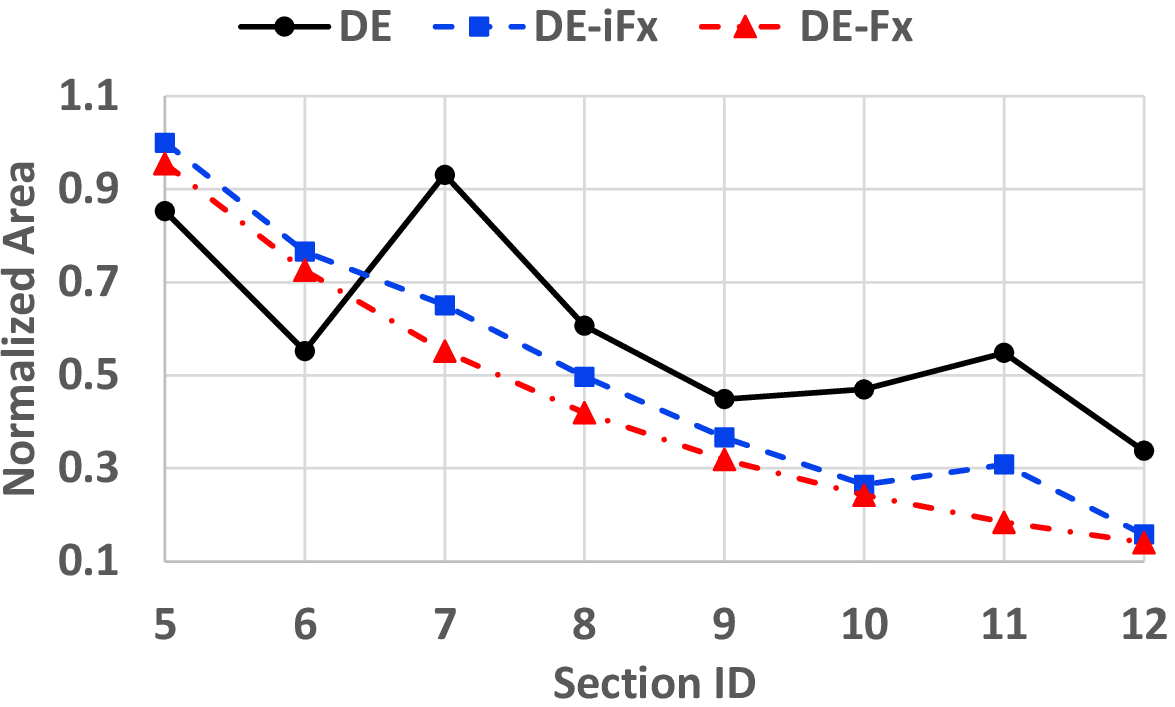,width=4.3cm}\label{sfig:outde3b24s}}\;
           \subfigure[Inner column using DE ]{\psfig{figure=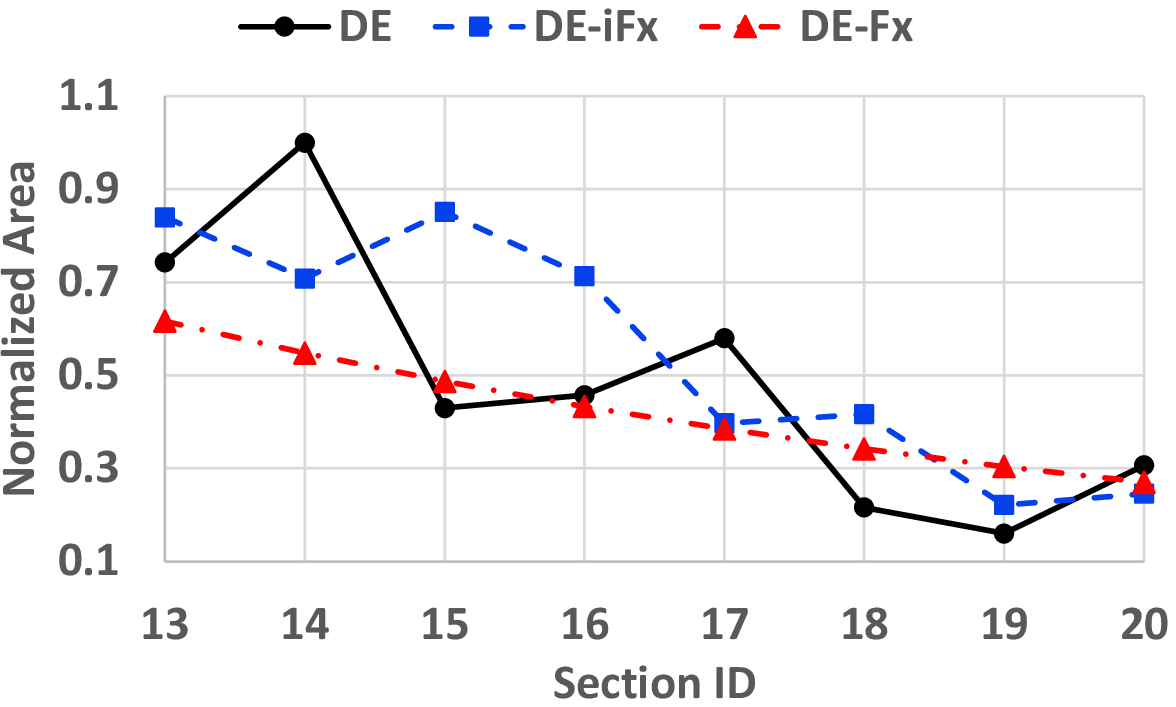,width=4.3cm}\label{sfig:inde3b24s}}\;}
\caption{Optimum design of columns cross-section areas for the 3-bay 24-story frame structure.}
\label{fig:section3b24s}
\end{figure}

The normalized outer and inner column cross-section areas of the best solutions using PSO and DE algorithms are presented in Figure~\ref{fig:section3b24s}.
As it is discussed (section 4.2), a column’s cross-section areas should be reduced with increasing height based on physics of the problem and engineering knowledge.
From Figure~\ref{sfig:outpso3b24s}, it can be seen that only PSO-$Fx$ has the expected trend in the results for the outer column.
Figure~\ref{sfig:inpso3b24s} shows that the optimum design of PSO without the proposed approach does not have the desired trend in the inner column cross-section areas, and the final design of PSO-$iFx$  slightly violate this condition.
However, for PSO-$Fx$ it is not the case, since the expected trend is considered by the variable functioning. In other words, all the solutions are forced to satisfy the condition and, therefore, the algorithms only search the solutions that have the trend and monotonically decrease with increasing height.
From Figures~\ref{sfig:outde3b24s} and~\ref{sfig:inde3b24s}, DE and DE-$iFx$ have the construction condition in neither outer nor inner columns.
And therefore, the DE-$Fx$ results are even more practical, as they satisfy the conditions.

\section{Conclusion and Discussions}
\label{sec:conclusions}

In this study, a concept-based approach is proposed to replace sets of variables with their functions by incorporating different sources of information.
For the steel frame design optimization problems, first the structure of the variables (cross-section areas) are explored using a grouping method and then a function is defined to relate each column cross-section areas by means of their heights.
A general equation is developed to relate a column cross-section areas by using a differential grouping and engineering points of view.
Here, the proposed approach is coupled with two well-known global optimization algorithms and its implementation is illustrated by a 50-stepped column design problem.
Design of three complex steel frame structures with different numbers of bays and stories are optimized as numerical case studies.
The proposed approach is coupled with optimization algorithms using two different strategies:
\begin{inparaenum}
\item Only for initialization; and
\item During the entire iterations.
\end{inparaenum}
From the results, it can be seen that using the proposed method as initialization can improve the results, and using the proposed approach during the iterations can significantly improve the convergence rates and final solutions, especially for the tall and more complex frame design optimization problems.
Using the proposed approach during initialization is significantly effective in all cases and improves the results for the studied steel frame structures.
Therefore, it can be used for seeding of frame design optimization problem, instead of randomly generating the initial solutions.
Although the relationship proposed for the column sections may not be exact, it achieves competitive results, and most of the time it produces better results in comparison with an algorithm that does not use the proposed approach.
In terms of numbers of function evaluations, if the proposed approach is used during the entire search process, it can significantly improve the convergence rate and reduce the number of required finite element analyses (which is time-consuming for real-world problems).
It should be noted that the proposed approach also considers the practicality aspect of a column cross-section areas which may not be achieved with other strategies.
The proposed function is applied to columns of moment resisting steel frame and, therefore, future research could focus on applying and modifying it to different type of frame systems~\cite{camp2013co} and real-world structures~\cite{azad2021design}.

\begin{appendix}
\section*{Appendix}
    \setcounter{table}{0}
    \renewcommand{\thetable}{A.\arabic{table}}

\begin{table}[htb]
\centering
\caption{Summary of notations used in this study}
\begin{tabular}{ll}
\toprule
Symbol / Acronym & Definition / Description \\
\midrule
$A$                & Cross-section area                                                            \\
$E$                & Elastic modulus                                                               \\
$H$                & Height of the frame structure                                                 \\
$K$                & Effective length factors of members                                           \\
$L$                & Length                                                                        \\
$M_{\mathrm{nx}}$  & Required nominal flexural strengths in the x-direction                        \\
$M_{\mathrm{ny}}$  & Required nominal flexural strengths in the y-direction                        \\
$M_{\mathrm{ux}}$  & Required flexural strengths in the x-direction                                \\
$M_{\mathrm{uy}}$  & Required flexural strengths in the y-direction                                \\
$P$                & Force                                                                         \\
$P_n$              & The nominal axial strength (tension or compression)                           \\
$P_u$              & The required strength (tension or compression)                                \\
$R$                & Maximum drift index                                                           \\
$RI$               & Inter-story drift index permitted by standard design code                     \\
$\mathbf{V}$       & Velocity component matrix                                                     \\
$W$                & Frame weight                                                                  \\
$\vec{x}$          & Design variable                                                               \\
$d$                & Inter-story drift                                                             \\
$f$                & Objective function                                                            \\
$f_{\mathrm{max}}$ & Objective function value of the worst feasible solution in the population     \\
$g$                & Constraint function                                                           \\
$g_{\mathrm{max}}$ & The largest known violation of a constraint                                   \\
$h$                & Height from the base (or ground)                                              \\
$h_j$              & Height of $j$th floor                                                         \\
$l$                & Length of each segment                                                        \\
$n_{\mathrm{c}}$   & Number of constraints                                                         \\
$n_{\mathrm{g}}$   & Number of groups                                                              \\
$n_{\mathrm{m}}$   & Number of members                                                             \\
$ns$               & Total number of stories                                                       \\
$r$                & Radius                                                                        \\
$\Delta_T$         & The maximum lateral displacement                                              \\
$\Omega$           & Search space                                                                  \\
$\phi_b$           & Flexural resistance reduction factor                                          \\
$\phi_c$           & Resistance factor (tension or compression)                                    \\
$\rho$             & Material density                                                              \\
$\sigma_i$         & Maximum stress in the $i$th member                                            \\
$\sigma_i^a$       & Allowable stress in the $i$th member                                          \\
$\zeta$            & A multivariate mathematical function from $\mathbb{R}^m$ to $\mathbb{R}^q$    \\
\bottomrule
\end{tabular}
\end{table}

\begin{table}[htb]
\centering
\caption{Summary of acronyms used in this study}
\begin{tabular}{ll}
\toprule
Acronym & Description \\
\midrule
AISC               & American Institute of Steel Construction                                      \\
DE                 & Differential evolution                                                        \\
DG                 & Differential grouping                                                         \\
DG2                & Extended differential grouping                                                \\
FE                 & Finite element                                                                \\
Fx                 & Variable functioning                                                          \\
LRFD               & Load and Resistance Factor Design                                             \\
PSO                & Particle swarm optimization                                                   \\
iFx                & Initialization by variable functioning                                        \\
\bottomrule
\end{tabular}
\end{table}

\end{appendix}

\section*{Acknowledgement}


Authors are grateful to Associate Professor Saeid K Azad of the Atilim University for his constructive comments and fruitful suggestions.


%
\section*{Replication of results}
Optimization algorithm codes and data for replication can be provided up on request.

\section*{Conflict of interest}

The authors declare that they have no conflict of interest.

\bibliographystyle{spbasic}      
\bibliography{./bibliography.bib}   


\end{document}